\documentclass[useAMS,usenatbib]{mn2e}

\usepackage{epsfig}
\usepackage{myaasmacros}
\usepackage{amsmath}
\usepackage{amssymb}
\usepackage{amsfonts}
\usepackage{subfig}
\usepackage{comment}

\title[Metal enrichment and galactic winds in cosmological zoom simulations]
{The effect of metal enrichment and galactic winds
  on galaxy formation in cosmological zoom simulations}
\author[Hirschmann et al.]{Michaela Hirschmann$^{1}$\thanks{E-mail:
mhirsch@oats.inaf.it}, Thorsten Naab$^{2}$, Romeel Dav\'{e}$^{3,4}$, Benjamin
D. \newauthor Oppenheimer$^{5}$, Jeremiah P. Ostriker$^{6,7}$, Rachel
S. Somerville$^{8}$, Ludwig Oser$^{2,7}$, \newauthor  Reinhard
Genzel$^{9}$, Linda J. Tacconi$^{9}$, Natascha
M. F\"orster-Schreiber$^{9}$, \newauthor Andreas Burkert$^{9,10}$, Shy
Genel$^{11}$\\  
$^{1}$INAF - Astronomical Observatory of Trieste, via G.B. Tiepolo 11,
I-34143 Trieste, Italy\\
$^{2}$Max-Planck-Institut f\"ur Astrophysik,
Karl-Schwarzschild Strasse 1, D-85740 Garching, Germany\\
$^{3}$Astronomy Department, University of Arizona, Tucson, AZ 85721,
USA\\ 
$^{4}$South African Astronomical Observatories, Observatory, Cape Town
7925, South Africa\\ 
$^{5}$Leiden Observatory, Leiden University, PO Box 9513, 2300 RA
Leiden, the Netherlands\\ 
$^{6}$Department of Astrophysical Sciences, Princeton University,
Princeton, NJ 08544, USA\\ 
$^{7}$Department of Astronomy, Columbia University,
New York, NY 10027, USA\\ 
$^{8}$Department of Physics and Astronomy, Rutgers University, 136
Frelinghuysen Rd, Piscataway, NJ 08854, USA \\
$^{9}$Max-Planck-Institut f\"ur extraterrestrische Physik (MPE),
Giessenbachstr. 1, 85748 Garching, Germany\\ 
$^{10}$Universit\"ats Sternwarte M\"unchen, Scheinerstr.1, D-81679
M\"unchen, Germany\\ 
$^{11}$Harvard-–Smithsonian Center for Astrophysics, 60 Garden Street,
Cambridge, MA 02138, USA} 

\begin{document}

\date{Accepted ???. Received ??? in original form ???}

\pagerange{\pageref{firstpage}--\pageref{lastpage}} \pubyear{2002}

\maketitle

\label{firstpage}

\begin{abstract}
We investigate the differential effects of metal cooling and
  galactic stellar winds on the cosmological formation of individual
  galaxies with three sets of cosmological, hydrodynamical
  zoom simulations of 45 halos in the mass range $10^{11} <
  M_{\mathrm{halo}} < 10^{13} M_\odot$. Models including both galactic
  winds and metal cooling (i) suppress early star formation at z
  $\gtrsim$ 1 and predict reasonable star formation histories for
  galaxies in present day halos of $\lesssim 10^{12} M_{\odot}$, (ii)
  produce galaxies with high cold gas fractions (30 - 60 per cent) at
  high redshift, (iii) significantly reduce the galaxy formation
  efficiencies for halos ($M_{\mathrm{halo}} \lesssim 10^{12}
  M_\odot$) at all redshifts in   overall good agreement with recent
  observational data and constraints from abundance matching, (iv)
  result in high-redshift galaxies with reduced circular velocities in
  agreement with the observed Tully-Fisher relation at  $z \sim 2$ and
  (v) significantly increase the sizes of low-mass galaxies
  ($M_{\mathrm{stellar}} \lesssim 3 \times 10^{10} M_\odot$) at high
  redshift resulting in a weak size evolution --  a trend in agreement 
  with observations. However, the low redshift ($z < 0.5$) star
  formation rates of massive galaxies are higher than observed
  (up to ten times). No tested model predicts the observed size
  evolution for low-mass and high-mass galaxies  simultaneously.
  Without winds the sizes of low-mass galaxies evolve to rapidly, with
  winds the size evolution of massive galaxies is too weak. Due to the
  delayed onset of star formation in the wind models, the metal
  enrichment of gas and stars is delayed and agrees well with
  observational constraints. Metal cooling and stellar winds are both
  found to increase the ratio of in situ formed to accreted stars -
  the relative importance of dissipative vs. dissipationless
  assembly. For halo masses below $\sim 10^{12} M_\odot$, this is
  mainly caused by less stellar accretion and compares well to
  predictions from semi-analytical models, but differs from
  abundance matching models as the in situ formed fractions of stellar
  mass are still too low in the simulations. For higher masses,
  however, the fraction of in situ stars is over-predicted due to the
  unrealistically high star formation rates at low redshifts.       
\end{abstract}


\section{Introduction}\label{intro}

Simple models for galaxy formation assume that gas gets trapped in
potential wells of dark matter halos, cools and condenses in central
disc, forms stars and self-enriches with a critical galactic stellar
mass of the order of $10^{12}M_\odot$ so that halos with masses $\gg
10^{13} M_\odot$ fragment into many galaxies (e.g. \citealp{Rees77,
  White78}). In hierarchical structure formation scenarios, these
processes are accompanied by merger events and continuous gas
inflow. However, unaltered inflow and  cooling of gas would grossly
over-estimate the star formation within a galaxy and result in too
massive galaxies known as the 'over-cooling' problem
(e.g. \citealp{White91,Balogh01}).  

It is plausible that additional 'feedback' processes regulate galaxy
growth, i.e. galactic winds driven by supernova explosions or
radiation pressure from young massive stars (e.g. \citealp{Martin05,
  Rupke05,Murray05}). A number of different numerical models for
stellar feedback have been developed and implemented in simulations
(e.g. \citealp{DallaVecchia12, Stinson12,Dave11a, 
  Hopkins11, Governato10, Schaye10, DallaVecchia08, 
  Oppenheimer06} and references therein) partly resulting in much
better agreement of the galaxy properties with observational
constraints. Especially for massive galaxies, energy-driven and/or
momentum-driven feedback from 
accretion onto a central black hole can additionally alter galaxy
formation and alleviate the over-cooling problem
(e.g. \citealp{Croton06, Somerville08, Hopkins05, Springel05a,
  Ciotti10, Teyssier11, Puchwein12, Martizzi12, Choi12,
  Barai13}). Overall, such feedback processes are expected to impact
the evolution of the stellar galaxy masses, their metal and gaseous
content, the efficiency of converting baryons into stars and, as a
consequence, also the stellar mass assembly (in situ star formation
vs. accretion of stars) and the size evolution (\citealp{Khochfar06b,
  DeLucia06, Naab07, Naab09, Oser10, Hirschmann12, Porter12, Oser12,
  Hilz12, McCarthy12, Lackner12, Hilz13}). 

Also at higher redshifts observations have now revealed tight
correlations between stellar mass, gas mass, star formation rate and
metal content like the relation between stellar mass and star
formation rate (SFRs), often termed the 'main sequence for
star-forming galaxies' (\citealp{Noeske07,Daddi07, Elbaz07,
  Genzel10}). Another example is the relation of stellar mass with
gas-phase metallicity  (\citealp{Tremonti04, Maiolino08,
  Andrews12}). Even though the total gas content of a galaxy is
difficult to measure (because atomic, molecular and ionized phases
must be accounted for), there are strong indications for galaxies with
higher stellar masses to have lower gas fractions
(\citealp{Saintonge11a, Peeples11}). In general, all these relations
evolve with redshift and any successful theoretical framework has to
account for this evolution (\citealp{Tremonti04, Gallazzi05, 
  Thomas05, Erb06, Elbaz07, Daddi07, Salim07, Maiolino08, Tacconi10, 
  Peeples11, Tacconi13, Andrews12}). 

Despite of the progress in numerical modelling of galaxy formation,
the detailed physics of star formation and feedback processes cannot
be directly simulated on the relevant small scales but is incorporated
in an approximate fashion on sub-resolution scales - more or less
motivated by the physics of star formation and observed
phenomenology. In has been shown that models driving galactic winds,
which can be powered by stellar winds and UV photons from young stars
and supernova explosions, can greatly improve the agreement of
simulations with observations (\citealp{Kauffmann93, Springel03, 
  DeLucia04, Oppenheimer06, Stinson06, Governato07, Oppenheimer08,
  Murante10, Governato10, Cen11, Wiersma11, Genel12, Barai12,
  McCarthy12, Haas12, Governato12, Hopkins12, Kannan13, Agertz13,
  Stinson13, Aumer13, Hopkins13, Angles13}). Such outflows are now directly
observed in most star-forming galaxies up to $z \sim 3$
(\citealp{Martin05, Rupke05, Weiner09, Steidel10, Genzel11,
  Newman12}). Energy and/or momentum injection from massive stars can
be modelled in many different ways into the numerical schemes of
simulations. Pure thermal supernova feedback, just heating the
surrounding gas, is inefficient (e.g. \citealp{Katz92, Steinmetz95,
  Katz96, Hummels12}) as the dense star-forming gas has a high
cooling-rate and the energy can be radiated away very quickly before
it can be  deposited into the ISM driving a wind (see, however, for a
recent improvement \citealp{DallaVecchia12}).  This is mainly a
  result of low resolution: thermal feedback works well at high enough
  resolution (better than 10~pc, see e.g. \citealp{Ceverino09}).
To overcome this problem (for simulation with not high-enough
resolution), energy injection from star-bursts and/or supernovae has
been modelled in many different ways: in a kinetic form
(e.g. \citealp{Navarro93, Cen00, Springel03, Oppenheimer06,
  DallaVecchia08}), by turning off radiative cooling temporarily
(e.g. \citealp{Stinson06, Governato07, Governato10, Piontek11,
  Stinson13}) or by forcing energy injection into the hot and cold gas
phase separately (e.g. \citealp{Scannapieco06, Murante10}).   

\begin{table*}
\centering
\begin{tabular}{ | p{0.5cm} || p{0.8cm} p{0.8cm} p{0.8cm} p{1.2cm} |
    p{0.8cm} p{0.8cm} p{0.8cm} p{1.2cm} | p{0.8cm} p{0.8cm} p{0.8cm}
    p{1.2cm} |} 
\hline  & & {\bf{NoMW}} & & & & {\bf{MNoW}} & & & & {\bf{MW}} & & \\ \hline
{\bf{ID}} & $M_{\mathrm{vir}}$ & $R_{\mathrm{vir}}$ & $ M_{\mathrm{stellar}}
$ & $M_{\mathrm{gas}}$ & $M_{\mathrm{vir}}$ & $R_{\mathrm{vir}}$ & $ M_{\mathrm{stellar}}
$ & $M_{\mathrm{gas}}$ & $M_{\mathrm{vir}}$ & $R_{\mathrm{vir}}$ & $ M_{\mathrm{stellar}}
$ & $M_{\mathrm{gas}}$  \\ \hline \hline
0163 & 966.8 & 310 & 28.14 & 0.793 & 914.6 & 305 & 35.81 & 0.559 &
904.7 & 304 & 36.58 & 0.645 \\ 
0190 & 699.2 & 279 & 27.32 & 0.840 & 685.6 & 277 & 31.35 & 0.578 &
657.7 & 273 & 31.92 &  1.04 \\ 
0209 & 699.5 & 279 & 18.34 & 0.462 & 678.6 & 276 & 20.74 & 0.342 &
681.3 & 276 & 19.67 &  0.780 \\ 
0215 & 668.7 & 274 & 25.48 & 0.521 & 659.2 & 273& 28.20 & 0.420 &
662.1 & 274 & 26.92 &  0.671 \\ 
0224 & 654.2 & 272 & 20.32 & 0.419 & 621.4 & 268 & 22.52 & 0.414 &
640.3 & 271 & 24.56 &  0.656 \\ 
0227 & 705.9 & 279 & 27.22 & 0.848 & 700.0 & 279 & 30.33 & 0.455 &
695.4 & 278 & 30.90 &  0.680 \\ 
0259 & 592.6 & 264 & 17.38 & 0.451 & 525.2 & 253 & 19.95 & 0.353 &
555.2 & 258 & 17.80 &  0.618 \\ 
0290 & 570.0 & 260 & 18.27 & 0.548 & 544.4 & 256 & 20.03 & 0.398 &
546.7 & 257 & 20.65 &  0.780 \\ 
0300 & 498.9 & 249 & 17.58 & 0.550 & 495.1 & 248 & 17.98 & 0.456 &
504.4 & 250 & 17.01 &  0.877 \\ 
0305 & 453.6 & 241 & 12.56 & 0.428 & 463.9 & 243 & 14.69 & 0.355 &
465.6 & 243 & 20.37 & 0.623 \\ 
0329 & 466.5 & 243 & 17.55 & 0.689 & 450.4 & 241 & 20.74 & 0.351 &
462.0 & 243 & 19.98 &  1.38 \\ 
0380 & 443.3 & 239 & 14.37 & 0.617 & 441.5 & 239 & 17.09 & 0.615 &
442.4 & 239 & 14.99 &  0.480 \\
0408 & 332.5 & 217 & 13.77 & 1.502 & 338.0 & 219 & 15.62 & 0.382 &
326.7 & 216 & 15.69 &  0.294 \\ 
0443 & 362.9 & 224 & 17.23 & 1.082 & 371.0 & 226 & 18.54 & 0.455 &
381.1 & 228 & 17.46 &  0.705 \\  
0501 & 306.1 & 212 & 14.79 & 0.408 & 306.3 & 212 & 13.93 & 0.216 &
304.6 & 211 & 14.79 & 0.575 \\ 
0549 & 281.4 & 206 & 10.50 & 0.404 & 276.6 & 205 & 10.60 & 0.328 &
292.7 & 208 & 10.58 & 0.396 \\ 
0616 & 260.8 & 201 & 12.93 & 0.442 & 257.3 & 200 & 13.18 & 0.287 &
261.3 & 201 & 13.37 &  0.790 \\ 
0664 & 208.5 & 186 & 8.719 & 0.236 & 241.8 & 196 & 9.801 & 0.305 &
245.8 & 197 & 10.0 & 0.553 \\
0721 & 194.2 & 182 & 10.56 & 0.220 & 192.1 & 181 & 12.04 & 0.220 &
196.0 & 182 & 10.77 &  0.720 \\
0858 & 188.6 & 180 & 10.30 & 0.397 & 186.0 & 179 & 11.43 & 0.225 &
189.1 & 180 & 10.93 &  0.212 \\
0908 & 172.1 & 175 & 9.245 & 0.506 & 169.7 & 174 & 9.883 & 0.197 &
169.4 & 174 & 9.077 &  0.514 \\
0948 & 162.7 & 171 & 6.314 & 0.146 & 160.6 & 171 & 9.241 & 0.174 &
164.8 & 172 & 8.471 & 0.571 \\
0959 & 158.0 & 170 & 6.015 & 0.157 & 155.5 & 169 & 10.10 & 0.237 &
161.1 & 171 & 9.379 &  0.315 \\
0977 & 139.6 & 163 & 2.743 & 0.273 & 125.6 & 157 & 4.964 & 0.486 &
131.4 & 160 & 2.774 &  1.25 \\
1017 & 139.7 & 163 & 6.988 & 0.182 & 140.5 & 163 & 8.181 & 0.195 &
141.8 & 164 & 6.453 &  0.862 \\
1061 & 139.3 & 163 & 6.840 & 0.191 & 137.7 & 162 & 7.609& 0.411 &
141.4 & 164 & 7.375 &  0.672 \\
1071 & 146.5 & 165 & 8.451 & 0.185 & 142.9 & 164 & 9.028 & 0.155 &
148.3 & 166 & 10.27 & 0.470 \\
1091 & 147.0 & 166 & 8.195 & 0.256 & 143.4 & 164 & 9.187 & 0.245 &
149.1 & 166 & 8.006 &  0.180 \\
1167 & 127.6 & 158 & 7.873 & 0.451 & 125.3 & 157 & 9.025 & 0.249 &
124.3 & 157 & 7.461 & 0.176 \\
1192 & 104.7 & 148 & 5.267 & 0.125 & 103.2 & 147 & 5.727 & 0.124 &
112.0 & 151 & 5.189 & 0.257 \\
1196 & 136.2 & 161 & 8.019 & 0.565 & 132.3 & 160 & 8.973 & 0.303 &
132.3 & 160 & 7.772 &  0.256 \\
1646 & 95.39 & 143 & 5.577 & 0.173 & 95.22 & 149 & 6.290 & 0.134 &
93.03 & 142 & 6.250 &  0.401 \\
1859 & 81.44 & 293 & 9.634 & 0.444 & 81.37 & 293 & 7.507 & 0.341 &
81.68 & 293 & 7.084 &  0.344 \\
2283 & 54.07 & 119 & 2.222 & 0.202 & 79.67 & 135 & 2.398 & 0.144 &
82.87 & 137 & 2.205 & 0.439 \\
2665 & 50.07 & 119 & 3.427 & 0.090 & 53.89 & 119 & 3.799 & 0.092 &
56.10 & 120 & 3.710 &  0.148 \\
3431 & 33.61 & 144 & 3.136 & 0.113 & 33.72 & 145 & 3.262 & 0.177 &
33.22 & 144 & 2.161 & 0.962 \\
3852 & 41.50 & 109 & 3.467 & 0.174 & 40.64 & 108 & 3.571 & 0.232 &
21.81 & 87.7 & 0.292 &  0.387 \\
4234 & 40.50 & 108 & 2.591 & 0.261 & 39.15 & 107 & 2.826 & 0.231 &
40.49 & 108 & 0.732 &  0.513 \\
4323 & 33.65 & 101 & 2.592 & 0.194 & 34.21 & 102 & 2.750 & 0.204 &
34.79 & 102 & 1.963 & 0.660 \\
4662 & 30.40 & 98.0 & 2.406 & 0.139 & 30.17 & 97.7 & 2.288 & 0.171 &
31.10 & 98.7 & 1.009 & 0.571 \\
5014 & 28.05 & 136 & 2.837  & 0.376 & 28.18 & 136 & 2.924 & 0.542 &
28.46 & 137 & 0.579 &  0.361 \\
5988 & 25.07 & 91.9 & 2.203 & 0.193 & 24.82 & 91.6 & 2.222 & 0.260 &
25.74 & 92.7 & 0.784 &  0.552 \\
6622 & 20.82 & 86.4 & 1.389 & 0.258 & 20.76 & 86.3  & 1.340 & 0.284 &
43.18  & 110 & 0.665 &  0.252 \\
6782 & 22.10 & 88.1 & 2.225 & 0.140 & 22.33 & 88.4 & 2.131 & 0.207 &
22.93 & 89.2 & 0.373 &  0.315 \\
7748 & 17.24 & 81.1 & 1.686 & 0.115 & 17.52 & 81.6 & 1.637 & 0.150 &
17.51 & 81.5 & 0.506 &  0.490 \\
\end{tabular}
\caption{Halo ID, halo virial mass ($M_{\mathrm{vir}}$), virial radius
  ($R_{\mathrm{vir}}$), stellar mass ($M_{\mathrm{stellar}}$) and gas
  mass ($M_{\mathrm{gas}}$) of the \textit{central} galaxies for the
  three models (45 initial conditions each). All masses are in units of
$10^{10} h^{-1} M_\odot$ and the virial radius in in units of $h^{-1}
  \mathrm{kpc} $.}
\label{sim_tab}
\end{table*}

In this paper, we use a phenomenological feedback scheme presented by 
\citet{Oppenheimer06, Oppenheimer08, Finlator08,Oppenheimer09, Dave09, 
  Oppenheimer10} and  \citet{Dave11a} who implemented scalings
expected for momentum-driven winds (\citealp{Murray05, Zhang12}) in
large scale cosmological simulations with a typical spatial resolution
of a few kpc. It is assumed that the gas outflow rate (i.e. the mass loading)
is \textit{inversely} proportional to the stellar velocity dispersion,
suppressing star formation in smaller systems, and that the wind
velocity is \textit{directly} proportional to the velocity dispersion
of a galaxy. With such a scaling for driving winds, high outflow rates
and frequent gas re-accretion can be generated leading to a continuous
cycle of baryons between galaxies and their surrounding intergalactic
medium (IGM) and this cycle is found to impact many of the observed
galaxy and IGM properties. Although most galaxies in these simulations
were not resolved sufficiently well, these models were successful in
matching a wide range of global observational data of galaxies and
their ISM properties, although some significant discrepancies still
remain.  

In this study, we present three sets of higher resolution,
cosmological zoom simulations of 45 individual galaxy halos using a
modified version of Gadget2 including a treatment for metal enrichment
(SNII, SNIa and AGB stars) and strong momentum-driven winds (as
described in detail in \citealp{Oppenheimer06, Oppenheimer08}). This
extends recent studies by \citet{Oser10, Oser12}, which our initial
conditions are based on, but containing only thermal SN feedback,
which is known to have only a weak effect on suppressing star
formation (see e.g. \citealp{Hirschmann12}). In the three simulation
sets, we add metal enrichment/cooling and the wind model separately to
investigate the respective impact on the conversion of gas into stars,
the in- and outflow rates of gas, the resulting cold gas fractions,
star formation rates and the metal content in the stellar and gaseous
phase in halos spanning a mass range of $10^{11} <  M_{\mathrm{halo}}
< 10^{13} M_\odot$ (Note that we also analyse the substructure within
a halo considering sub-halo masses down to even $10^{10}
M_\odot$). Additionally, we examine the effect on the bi-modality in
the stellar mass assembly, i.e. the contribution of in situ formed and
accreted stars to the overall stellar content and on galaxy scaling
relations, e.g. the mass-size relation or the Tully-Fisher
relation. \citet{Oser10} found that massive galaxies form their stars
early in time and grow mainly via accretion of stellar systems at
later times explaining the observed stellar downsizing for massive
galaxies. In addition, we demonstrate in this study that galactic
winds are essential to reproduce \textit{the observed stellar
  downsizing concerning low-mass galaxies}. Overall in this work, we
go beyond previous studies (\citealp{Finlator08, Dave09, 
  Dave11b}) by using the same simulation code, i.e. the same SN
feedback prescription, but achieving a significantly increased
resolution (by about 1-2 orders of magnitude) in gas/dark matter
particle mass.  This approach is still useful at higher resolution as
it allows to test for convergence of the models in predicting
different galaxy properties (see e.g. \citealp{Vogelsberger13}).
Second, our resolution is higher but still not high enough to reliably
apply more direct feedback models (e.g. the Gasoline model requires an
at least one of order of magnitude better resolution compared to ours)
and it is far from having a resolution where physical processes
driving winds can be explicitly captured. Third, with higher
resolution we can investigate spatial information such as galaxy
sizes, the maximum circular velocities or the radial distributions of
metallicities (which will be the topic of a follow-up study).

Despite of the recent progress in numerically modelling the
hydrodynamics using e.g. a moving mesh algorithm (e.g.
\citealp{Springel10, Sijacki12, Keres12, Vogelsberger13}), the
implementations for star formation and stellar feedback still dominate
the end results of how galaxies form and evolve, rather than the
details of the implementation of the hydrodynamics.
\citet{Scannapieco12}, for example, have explicitly demonstrated
that different numerical techniques typically lead to differences as
more gas seems able to cool and become available for star formation in
grid-based codes than in SPH. However, they emphasise that this effect
is small compared to the variations induced by different feedback
prescriptions. In a recent study of \citet{Marinacci13} they have
shown that to simulate ``realistic'' Milky-Way-like galaxies
(e.g. disk-dominated systems with flat rotation curves) using a moving
mesh code, also a (new) strong feedback model is necessary. Overall,
this justifies the SPH implementation of GADGET2 used in this paper
and demonstrates the importance of studying the effect of galactic
winds irrespectively of the used numerical implementation. 

\begin{figure*}
\begin{minipage}[b]{1.0\linewidth}
  \epsfig{file=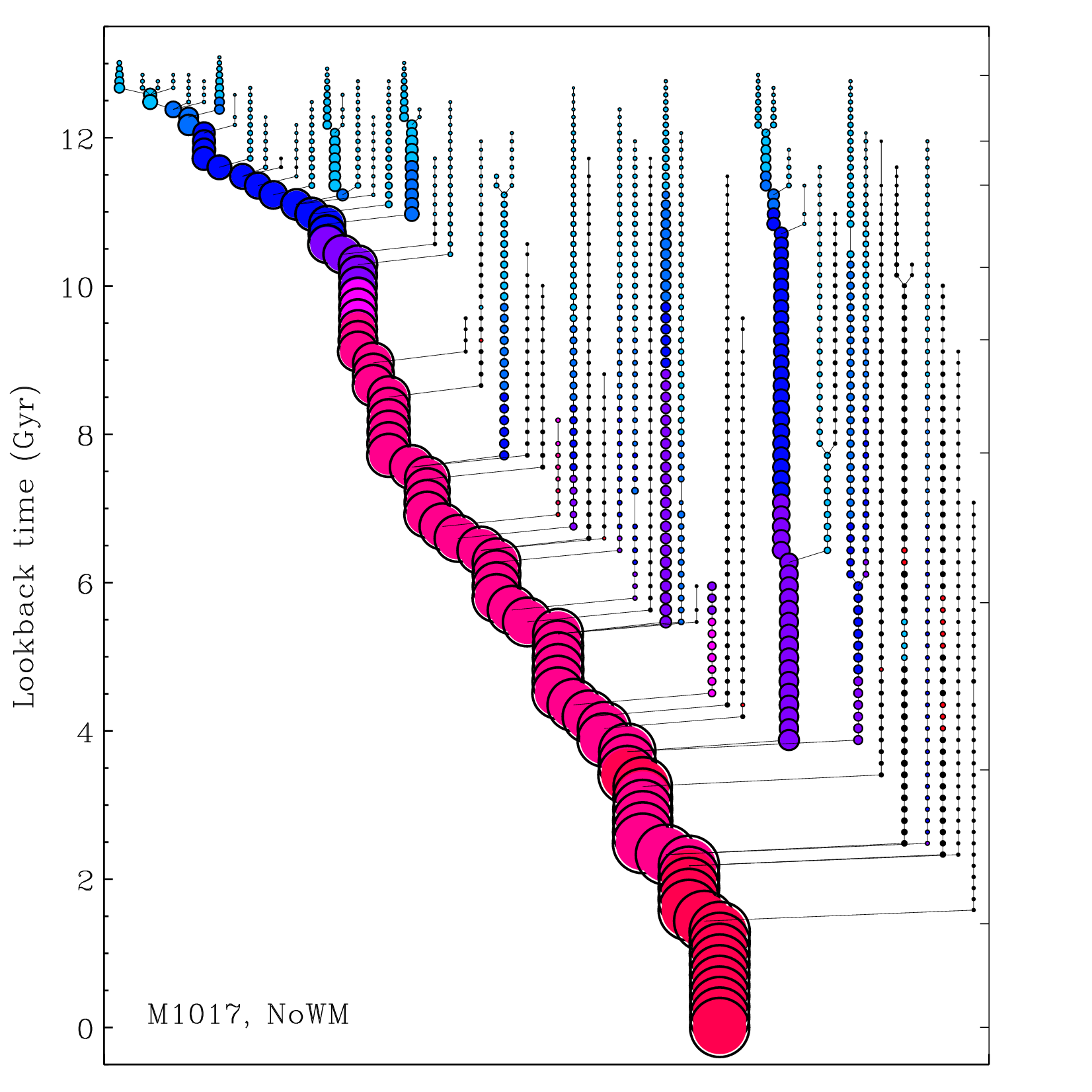,
    width=0.36\textwidth}\hspace{-1.3cm}
\epsfig{file=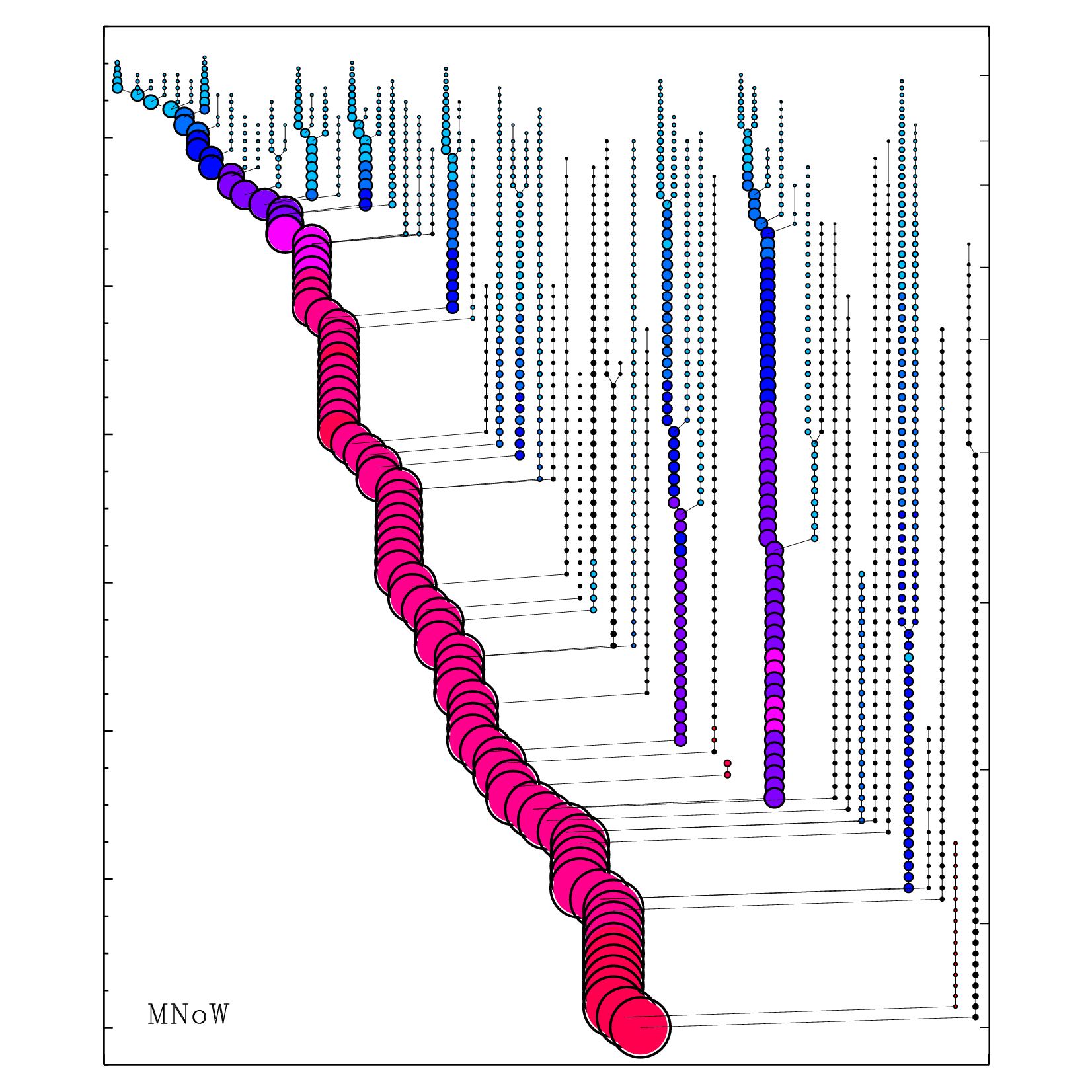,
    width=0.36\textwidth}\hspace{-1.3cm}
 \epsfig{file=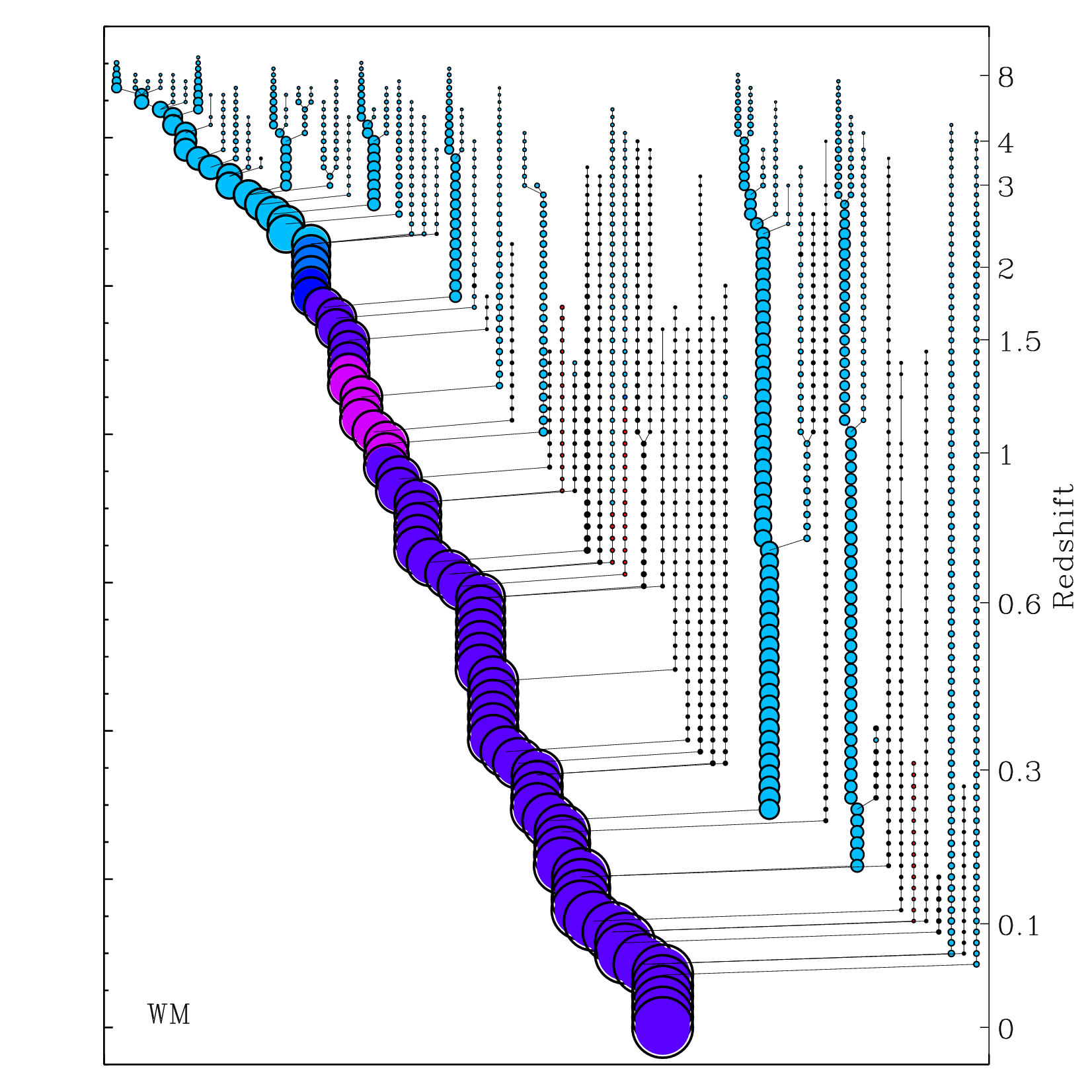,
    width=0.36\textwidth}
  \epsfig{file=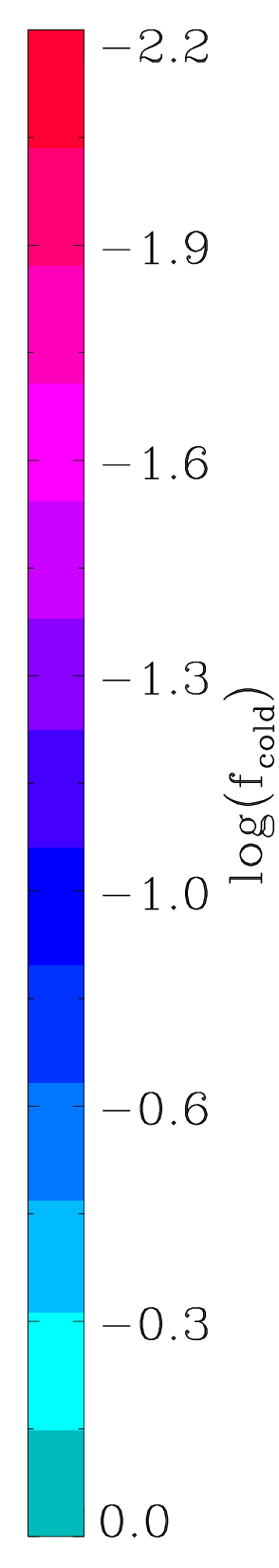,width=0.064\textwidth} 
\end{minipage}
  \caption{Three merger trees for halo 1017 with a mass of $\approx
    10^{12} M_{\odot}\ h^{-1}$ for a hydrodynamical zoom-simulation
    without metal enrichment and SN winds (NoWM, left panel), with metal
    enrichment but no SN winds (MNoW, middle panel) and including both
    metal enrichment and SN winds (WM, right panel). Black circles
    indicate the dark matter halos (as identified by the halo finder)
    and the size of the circles is scaled by the square root of the
    dark matter virial radius. In addition, the cold gas fractions
    $f_{\mathrm{cold}}$ of the galaxies are colour-coded. The merger
    trees visualise that simulations including strong winds
    significantly increase the cold gas fraction (bluer colour).}  
 {\label{Mergertrees}}
\end{figure*}

The paper is organised as follows. Section \ref{model} provides an
introduction into our simulation set-up and the construction of the 
merger trees. In section \ref{SFRcold} we discuss the effect of
galactic winds and metal cooling on star formation rates, in- and
outflow rates, cold gas fractions and the baryon conversion
efficiencies. We continue to focus on the bi-modality of star
formation (in situ formed versus accreted stellar content) in our
different simulation sets in section \ref{AccIns} and on galaxy
scaling relations in section \ref{Size}. In addition, in section
\ref{Metal} we present stellar and gaseous metallicities and ages of
our simulated galaxies. A final summary and discussion of this work is
given in section \ref{Discussion}.

\section{High-resolution simulations of individual galaxy halos}\label{model} 

\subsection{Simulation setup}\label{setup}

The cosmological zoom simulations presented in this paper are based on
the initial conditions described in detail in \citet{Oser10, Oser12}.
We briefly review the simulation setup here, but refer the reader to
the original papers for more details. The dark matter halos for
further refinement were selected from a dark matter only N-body
simulation with a co-moving periodic box length of $L=100\
\mathrm{Mpc}$ and $512^3$ particles (\citealp{Moster10}). We assume a
$\Lambda$CDM cosmology based on the WMAP3 measurements (see
e.g. \citealp{Spergel03}) with $\sigma_8 = 0.77$, $\Omega_{m}=0.26$,
$\Omega_{\Lambda}=0.74$, and $h=H_0/(100\ \mathrm
{kms}^{-1})=0.72$. The simulation was started at $z=43$ and run to
$z=0$ with a fixed co-moving softening length of $2.52\ h^{-1}
\mathrm{kpc}$ and a dark matter particle mass of $M_{\mathrm{DM}} = 2
\times 10^8 M_{\odot}\ h^{-1}$. Starting at an expansion factor of 
$a=0.06$ we constructed halo catalogues for 94 snapshots until $a=1$
separated by $\Delta a =0.01$ in time. From this simulation, we picked
$45$ halos identified with the halo finder algorithm $FOF$ at 
$z=0$. To construct the high-resolution initial conditions for the
re-simulations, we traced back in time all particles closer than $2
\times r_{200}$ to the centre of the halo in any snapshot and 
replaced them with dark matter as well as gas particles at higher
resolution ($\Omega_b=0.044, \Omega_{DM}=0.216$). In the high
resolution region the dark matter particles have a mass resolution of
$m_{\mathrm{dm}} = 2.5\cdot 10^7 M_{\odot}h^{-1}$, which is 8 times
higher than in the original simulation, and the gas particle masses
are $m_{\mathrm{gas}} = m_{\mathrm{star}} = 4.2\cdot 10^6
M_{\odot}h^{-1}$. The co-moving gravitational softening length for the
gas and star particles is $400\ h^{-1} \mathrm{pc}$ and $890\ h^{-1}
\mathrm{pc}$ for the high-resolution dark matter particles. For two
halos, we have also performed simulations with 16 times higher mass
resolution than in the original DM-only simulation (see appendix
\ref{resolution}).  

To model the gas component we use the entropy conserving formulation
of SPH (\textsc{Gadget}-2, \citealp{Springel05}) with the extension of
\citet{Oppenheimer06, Oppenheimer08} including a prescription for
metal enrichment and momentum-driven winds. This version includes
ionisation and heating by a spatially uniform, redshift dependent
background radiation according to \citet{Haardt01}, where
re-ionisation takes place at $z \approx 6$ and the radiation field
peaks at $z \approx 2-3$. Gas particles undergo radiative cooling down
to $10^4$K  under the assumption of ionisation equilibrium; we account
for metal-line cooling using the collisional ionisation equilibrium
tables of \citet{Sutherland93}. Following \citet{Springel03}, stars
are formed from dense gas clouds using a sub-resolution multi-phase
model which describes condensation and evaporation in the interstellar
medium (\citealp{McKee77}). We have a density threshold for star
formation of $n_{th} = 0.13\ \mathrm{cm}^{-3}$, which is calculated
self-consistently in a way that the equation of state is continuous at
the onset of star formation. This value is the same as in the work of
\citet{Oppenheimer08} and \citet{Dave11a}, but lower than in
\citealp{Oser10} (who have a value of $n_{th} = 0.23\
\mathrm{cm}^{-3}$). Besides, we adopt a \citet{Chabrier03} IMF
throughout this study implying a fraction of stars, which results in
type II supernovae, of $f_{SN} = 0.198$ (different to
\citealp{Oser10}, who assume a Salpeter IMF with $f_{SN} =  0.1$). The
model is tuned via a single parameter, the star formation rate
timescale, using simulations of isolated disk galaxies to reproduce
the observed Schmidt-Kennicut relation. Note that in cosmological
simulations the result may, however, deviate from the observed
relation (see \citealp{Hirschmann12}).   

Following \citet{Oppenheimer08}, we account for metal enrichment from
supernovae type II (SNII), type Ia (SNIa) and asymptotic giant branch
(AGB) stars and four elements (C, O, Si, Fe) are tracked individually.
The SNII enrichment follows \citet{Springel03} using an instantaneous
recycling approximation, but is modified by adopting metallicity
dependent yields from the nucleosynthesis calculations by
\citet{Limongi05}. The SNIa rate is modelled following the
two-component parametrisation from \citet{Scannapieco05}, where one
component is proportional to the stellar mass (slow, delayed
component) and the other to the SFRs (rapid component). Besides the
production of metals, each SNIa is assumed to deposit $10^{51}$ ergs of
energy, and this energy is added thermally directly to the
gas particle. AGB stars mainly provide feedback in form of mass
(energy can be neglected as most mass leaves the AGB stars with
velocities far below $100$ km/s) and produce carbon and oxygen (while
silicon and iron remains almost unprocessed). To determine the stellar
mass loss rate from non-SN stars as a function of age and
metallicity we use the \citet{Bruzual03} stellar synthesis models and
interpolate in age and metallicity.    

The momentum-driven wind model is based on the wind model of
\citet{Springel03}: outflows are directly tied to the star formation
rates using the relation $\dot{M}_{\mathrm{wind}} = \eta
\dot{M}_{\mathrm{SF}}$, where $\eta$ is defined as the mass loading 
factor. Star forming gas particles get stochastically kicked
vertically to the disc and are thus,
launched in a wind with the probability $\eta$. A selected wind
particle is given the additional velocity of $v_w$ in the direction of
$\mathbf{v} \times \mathbf{a}$, where $\mathbf{v}$ and $\mathbf{a}$
are the velocity and acceleration of a particle, respectively
(\citealp{Springel03}). Subsequently, the gas particles are decoupled
from hydrodynamics for a short time in order to escape their dense,
star-forming regions and eventually to leave their galaxies (see
e.g. \citealp{DallaVecchia08}). These particles are only allowed to
again interact hydrodynamically as soon as they either reach a SPH
density less than 10 per cent of the SF density threshold or the time
it takes to travel 30~kpc at the wind velocity $v_w$ (note that the
first case significantly exceeds the instances of the second
case). The values of $\eta$ and $v_w$ define the wind model: while
\citet{Springel03} used constant values for these parameters,
\citet{Oppenheimer06} adopt a momentum-driven wind model and introduce
a scaling with the velocity dispersion of the galaxies motivated by
observations of galactic super-winds of \citet{Martin05, Rupke05} and
by analytical calculations of \citet{Murray05, Zhang12}. The magnitude
of the kick is given by   
\begin{equation}
v_{\mathrm{w}} = \sigma(2.9+4.29\sqrt{f_{\mathrm{L}}-1}),
\end{equation}
where $f_{\mathrm{L}}$ is the luminosity factor (critical luminosity
in terms of Eddington luminosity to expel gas out of the
galaxy). $f_{\mathrm{L}}$ is randomly selected from $f_{\mathrm{L}} =
[1.05-2]$ for each particle following the observations by
\citet{Rupke05} and is dependent on metallicity due to a higher UV 
photon output from lower metallicity stellar populations
\begin{equation}
f_{\mathrm{L}} = f_{\mathrm{L},\odot} \times 10^{-0.0029\times
  (\log Z +9)^{2.5} + 0.4177}.
\end{equation}
The mass-loading factor $\eta$ (i.e. the fraction of star-forming
particles, which get kicked) is calculated according to
\begin{equation}\label{massload}
\frac{\dot{M}_{\mathrm{w}}}{SFR} = \eta = \frac{\sigma_0}{\sigma},
\end{equation}
where $\sigma \propto M_{\mathrm{gal}}^{1/3}$ is the velocity
dispersion which is calculated from the galaxy mass using an
on-the-fly group finder. $\sigma_0$ is a constant which is set to
reproduce the overall evolution of the SFR density in
\citet{Oppenheimer06} (here: $\sigma_0 = 300\ \mathrm{km/s}$). For low
resolution simulations of large cosmological boxes this empirical
model was shown to reproduce the observed metal enrichment of the
intergalactic medium at $2<z<5$ (\citealp{Oppenheimer06}), the mass
metallicity relation (\citealp{Finlator08}), the cold gas fractions
(\citealp{Dave11b}), the SFRs (\citealp{Dave11a}) and baryon
conversion efficiencies (\citealp{Dave09}) at various redshifts. 
This model is clearly empirical and phenomenological. However, the
implemented scalings are consistent with observations and analytical
estimates for momentum-driven winds. This model captures the effect of
the wind but not its origin on unresolved scales. In this sense it
might be as valuable as many other sub-resolution models. The model
itself regulates how much gas is expelled from a galaxy with a given
SFR and mass. The mass loading regulates the delayed formation of
galaxies (the star formation histories, see section \ref{SFR}) what
might therefore be considered as a model input. Other galaxy
properties, like the size evolution or the fractions of in situ to
accreted stellar mass, the evolution of the mass-metallicity relation
can be considered as model predictions. 

\begin{figure*}
\begin{center}
  \epsfig{file=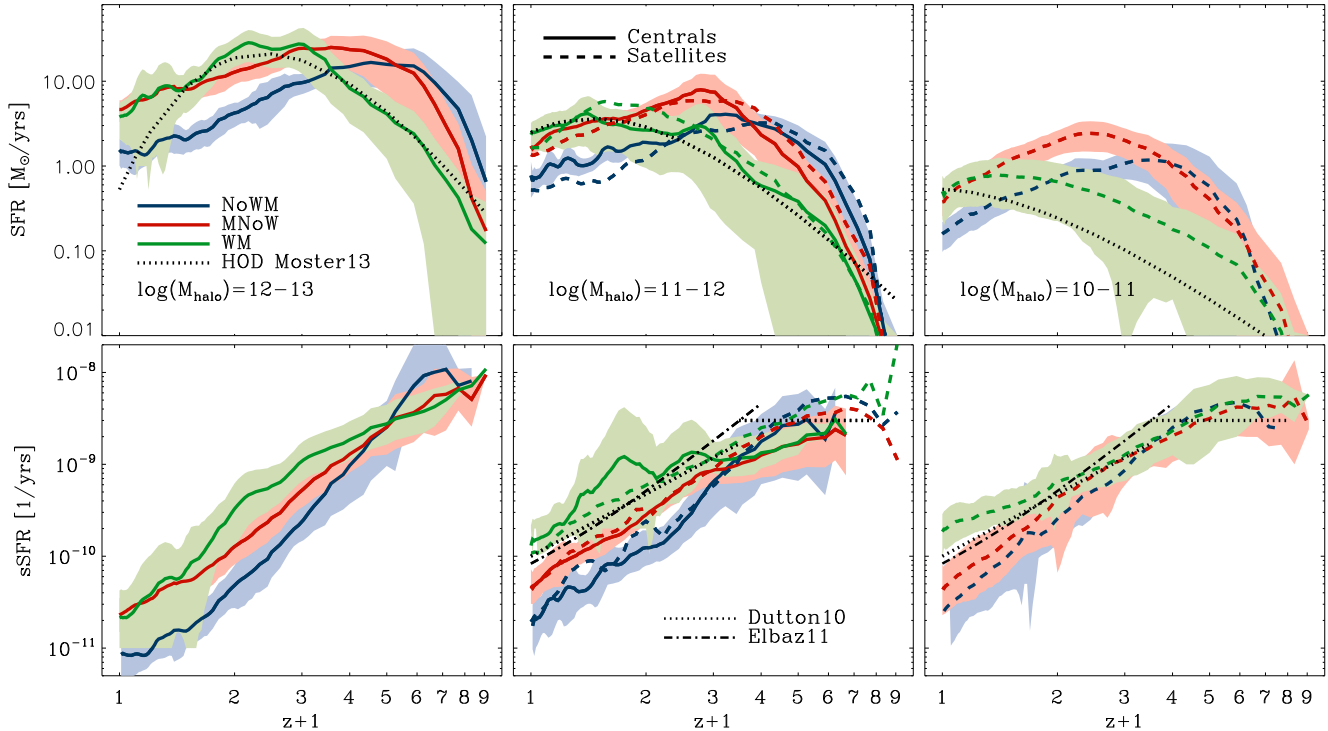, width=01.0\textwidth}
  \caption{Top row: Mean star formation rates (SFR) versus redshift
    (z+1) for the three models (blue: NoWM, red: MNoW, green: WM) in
    three different halo mass bins (different columns, considering the
    \textit{present-day} halo masses). Centrals are illustrated by
    solid, satellites by dashed lines. The coloured areas indicate the
    1-$\sigma$ scatter of the mean SFRs. The early star formation on
    the WM runs is efficiently by galactic winds and is in reasonable
    agreement with SHAM predictions (\citealp{Moster12}, black dotted
    lines) for halo masses above $10^{11}M_\odot$. Bottom row: Same as
  in the top row, but for the mean specific star formation rates
  (sSFR's). The WM runs can reproduce the decline of the observed
    sSFR's (of star-forming galaxies) with redshift: $\propto
    (z+1)^{2.25}$ for galaxy masses between $10^9-10^{11} M_\odot$
    (illustrated by black dashed lines, see observational compilation
    of \citealp{Dutton10,Reddy12}) and the one found by
    \citet{Elbaz11} (black dotted-dashed lines).}  
 {\label{SFR_evol}}
\end{center}
\end{figure*}

To disentangle the effects of metal enrichment and galactic winds we
study the influence on galaxy properties separately by running the
full set of simulations with three different models: 
 
\begin{itemize}
\item{{\bf{NoWM}}: No metal enrichment (only gas cooling for a
    primordial composition of hydrogen and helium) and no galactic
    winds (only thermal energy injection due to SNII explosions),
    comparable to the results of \citet{Oser10}, \citet{Oser12} and
    \citet{Hirschmann12}.} 
\item{{\bf{MNoW}}: Metal enrichment and metal cooling, but no
    galactic winds}
\item{{\bf{WM}}: Metal enrichment and momentum-driven
    galactic winds}
\end{itemize}

Table \ref{sim_tab} summarises global properties (halo mass, virial
radius, stellar mass and gas mass) of the 135 re-simulations. 
Note that all simulations include thermal supernova feedback as
described in detail in \citet{Springel03}\footnote{A certain fraction
of  formed stars is expected to explode immediately as supernovae
which heat the surrounding gas with an energy input of
$10^{51}$~ergs.}. For our analysis, we will not only consider the
central galaxy within the re-simulated halo, but also the
corresponding satellite galaxies with halo masses above $10^{10}
M_\odot$ within the virial radius of the parent halo at $z=0$. The halo
masses of satellites are determined at the time of their accretion
onto their final parent halos. Including 
satellite galaxies in our analysis extends our galaxy sample towards
lower masses which are stronger affected by the wind feedback.

\subsection{Merger trees}\label{trees}

We extract the merger trees for the dark matter component directly
from the cosmological re-simulations as described in
\citet{Hirschmann10}. For every snapshot at a given redshift, we first
identify individual dark matter haloes using a FOF
(Friends-of-Friends) algorithm with a linking length of $b=0.2$ 
(\citealp{Davis85}). In a second step we extract the sub-halos of every
FOF group using the \textsc{Subfind} algorithm
\citep{Springel01GAD}. This halo finder identifies over-dense regions
and removes gravitationally unbound particles. In this way we split
the FOF group into a main or host halo and its satellite halos. In
most cases, $90\%$ of the total mass is located in the main halo.

The sizes and virial masses of the main halos (i.e. the most massive
\textsc{Subfind} halos) are determined by a spherical over-density
criterion. The minimum halo mass is set to 20 particles ($5 \times
10^8 M_{\odot}/h$). The mass of a central object is defined by the dark
matter mass within the virial radius using the over-density
approximation in the spherical collapse model according to
\citet{Bryan98}. The algorithm to connect the dark matter halos
between the snapshots at different redshifts is described in detail in
\citet{Maulbetsch07}. The branches of the trees for $z=0$ halos are   
constructed by connecting the halos to their most massive progenitors
(MMP)  at previous snapshots. Thereby, halo $j$ with $n_j$ particles
at redshift $z_j$ with the maximum probability $p(i,j)$  is chosen to
be a MMP of halo $i$ containing $n_i$ particles at redshift $z_i$
(where $z_j < z_i$). The probability $p(i,j)$ is defined as   
\begin{align}
p(i,j) & = \frac{n_{ov}(i,j)}{n_{max}(i,j)} \quad \text{with} \\
n_{ov} & = n_i(z_i) \cap n_j(z_j) \quad \text{and} \nonumber\\
n_{max}(i,j) & = \max(n_i(z_i),n_j(z_j)) \nonumber
\end{align}
Here, $n_{ov}$ is the number of particles found in both halos and
$n_{max}$ is the particle number of the larger halo. We remove `fake'
haloes which exist only within one time step and have no connection to 
any branch (halo masses are generally near to the resolution
limit). The low redshift ends of the branches are then checked for
mergers. A halo $j$ at $t_j$ is assumed to merge into halo $i$ at
$t_i$, if at least $50 \% $ of the particles of halo $j$ are found in
halo $i$. In case of a merger the branches are connected.  

\begin{figure}
\begin{center}
  \epsfig{file=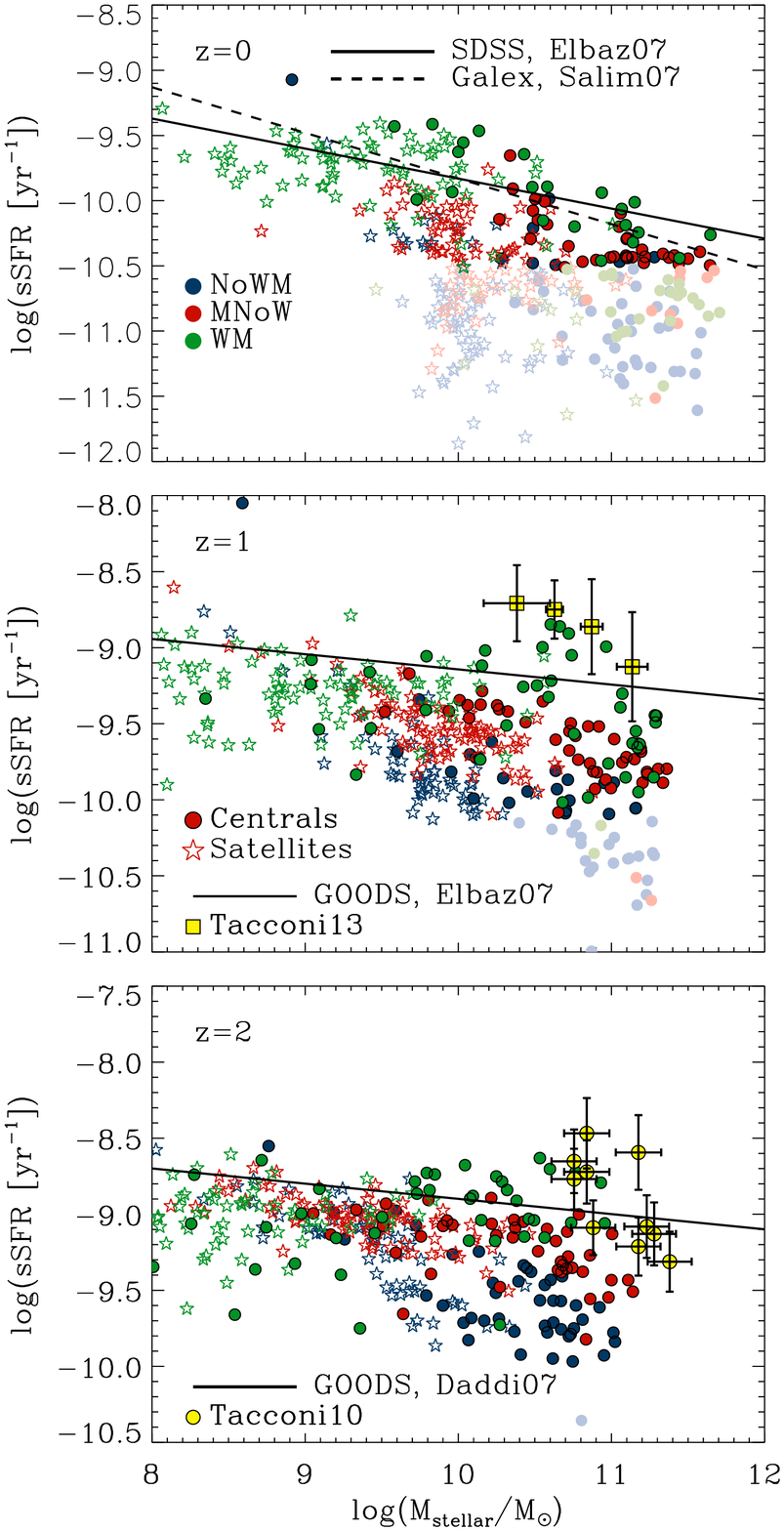, width=0.4\textwidth}
 \caption{sSFR's versus galaxy masses at z=0,1,2 (top, middle, and 
   bottom panel) for the three models for \textit{star-forming}
   central/satellite galaxies (with $\mathrm{sSFR} >
   0.3/t_{\mathrm{Hubble}}$, bright circles/stars) and \textit{quiescent}
   central/satellite galaxies (with $\mathrm{sSFR} <
   0.3/t_{\mathrm{Hubble}}$, light circles/stars), respectively. The
   simulation results are compared to the observed star-forming
   sequence of galaxies from SDSS and GOODS (black solid lines,
   \citealp{Elbaz07, Daddi07}) and to recent observational data
   from \citet{Tacconi10} and \citet{Tacconi13} (yellow symbols). MW
   runs reproduce the observed relation fairly well at all redshifts.
 } 
{\label{sSFR_Mgal}}
\end{center}
\end{figure}

Note that the tree-algorithm is only applied to the dark matter
particles -- star or gas particles are not separately traced back in
time. They are assumed to follow the evolution of the dark matter.
Therefore, we assign to each dark matter halo in a tree a hot/cold
phase gas mass by counting hot/cold gas particles within the virial
radius of the central halo. The stellar and cold gas particles within
$1/10$ of the virial radius (:$=r_{10}$) are defined as the stellar
and gas mass of the central galaxy. We distinguish between hot and
cold gas particles by using the following definition:   
\begin{eqnarray}\label{coldhot}
\log T < 0.3 \log\rho + 3.2\ \rightarrow \mathrm{cold}\\
\log T > 0.3 \log\rho +3.2\ \rightarrow \mathrm{hot},
\end{eqnarray}
where $T$ is the temperature and $\rho$ is the density divided by the
mean baryon density. Note that the above discrimination between hot
and cold gas was established by looking directly at the phase diagrams
of the re-simulations, where we have divided between the gas in the
disk heated by SN feedback and the shock heated gas. With the above
definition for cold gas we mainly capture the dense, star-forming
gas. 

In Fig. \ref{Mergertrees} we visualise three merger trees of the same
re-simulated halo with a virial mass of $1 \times 10^{12}M_{\odot}$
(M1071) for the three different simulations sets. The size of the
black circles approximates the dark matter halo mass as they scale
with the square root of mass normalised to the final dark matter halo
mass. The cold gas fractions ($=M_{\mathrm{cold}} / (M_{\mathrm{cold}}
+ M_{\mathrm{stellar}})$) of the galaxies residing in the dark matter
halos are colour-coded (as indicated by the colour bar). In all three
models, galaxies at high redshift contain a larger cold gas content,
which either turns into stars or is heated towards lower
redshifts. However, simulations including strong winds (WM) reveal a
significantly larger cold gas fraction than the simulations without
winds, at all redshifts.

\section{Conversion of gas into stars}\label{SFRcold} 

An important and useful parameter describing galaxy formation is the 
efficiency with which gas is turned into stars. This efficiency might
vary with time and redshift and is influenced by gas outflows which
not only determine the cold gas mass and the stellar content in a
galaxy but also influence 'secondary' galaxy properties such as the
metal content, the  stellar mass assembly, i.e. the relative
contributions from in situ star formation and accretion of stars, and
as a consequence the structural evolution of galaxies over cosmic
time.   

\subsection{Star formation rates}\label{SFR}

We start by investigating the star formation rates (SFRs) of our
simulated galaxies within $r_{10} := 1/10 \times r_{\mathrm{vir}}$
(for the satellites we use $r_{10}$ of their own sub-halo), shown
  in the top row of Fig. \ref{SFR_evol}. We consider in our analysis
  both central (solid lines) and satellite galaxies (dashed lines) in
  the three different simulation sets (green: WM; red: MNoW; black:
NoWM). We have binned the centrals and satellites into overall three
halo mass bins: $12 < \log{M_{\mathrm{gal}}/M_{\odot}} < 13$ \&  $11 <
\log{M_{\mathrm{gal}}/ M_{\odot}} < 12$ for centrals and $11 <
\log{M_{\mathrm{gal}}/ M_{\odot}} < 12$ \& $10 < \log{M_{\mathrm{gal}}
  /   M_{\odot}} < 11$ for satellites (different panels) and averaged
over the SFRs in each mass bin. 

The NoWM runs show significant early star formation: for the most
massive halos, the SFRs peak at $z \sim 5$ and are decreasing 
towards lower redshifts (top left panel). With decreasing halo mass,
the peaks of the SFRs are slightly moved towards lower redshifts
(e.g. $z \sim 3$ for halos with $10 < \log{M_{\mathrm{gal}} /
  M_{\odot}} < 11$, bottom right panel), illustrating a mild
trend of stellar downsizing or anti-hierarchical stellar
evolution. According to \citet{Neistein06} this a ``natural''
consequence of CDM models. For the MNoW runs, the peak of star
formation is slightly shifted towards lower redshifts ($z=2-4$
depending on the halo mass bin), again followed by a decline. The
shift of the peak is due to a higher star formation in a larger number 
of small halos as a consequence of more efficient cooling (in low-mass 
halos). In other words, gas cools and turns into stars preferentially
in smaller halos rather than falling into larger ones at high
redshifts (and thus, delaying the peak of the SFR in larger
galaxies). Simulations with wind feedback (WM runs) reveal a more  
moderate increase of star formation at high redshifts. For the high
and intermediate mass bins, star formation peaks between $z=1-2$,
while for the low mass bin, the SFR peaks below $z=0.5$. Therefore,
the WM model predicts a much more pronounced trend of stellar
downsizing than the NoWM model: low mass galaxies form at
significantly later times than their corresponding (low-mass) host
halos which - in a cold dark matter dominated universe preferentially
form at early times (\citealp{White91}). The early formation of low
mass galaxies in the WM runs is suppressed due to the efficient
ejection of star forming gas (see equation \ref{massload}),
particularly in low mass galaxies and also at high redshifts (as here
the re-infall rate is still low in contrast to low redshifts where
recycled gas accretion becomes the dominant source of gas infall for
new star formation, see e.g. \citealp{Oppenheimer06, Oppenheimer10}).    

The time evolution of the SFRs is compared to predictions from the
sub-halo abundance matching (SHAM) model as presented in
\citet{Moster12} (dotted black lines in the top row of Fig.
\ref{SFR_evol}).  Note that we have used the average halo mass in
each mass bin as an input value for the SHAM fitting functions. For
halo masses above $10^{11} M_\odot$, we find a good agreement between
the WM simulations and the SHAM model. Only at low redshifts (below
$z=1$) the WM runs predict higher SFRs for massive galaxies compared
to the SHAM model as too much gas is cooling and forming stars. This
implies that in massive galaxies a further mechanism might be needed
to efficiently suppress star formation as e.g. feedback from accreting
black holes (\citealp{Croton06, Sijacki07, Somerville08, Puchwein08,
  Booth09, Fabjan10, McCarthy10, vandeVoort11, Fanidakis11,
  Bower12, Hirschmann12a, Dubois13, Puchwein13}). 

Turning to the lowest halo mass bin (top right panel), even the SFRs in
the WM simulations are slightly higher than the SHAM predictions for
the whole redshift range, although low-mass satellites match the
observed SFR - stellar mass relation well, as we will discuss later in
Fig. \ref{sSFR_Mgal}. This indicates that these low
mass halos contain too massive galaxies (see also baryon conversion
efficiencies in Fig. \ref{Conveff}). This over-production of stars at
early times in halos with masses between $10^{10} - 10^{11} M_\odot$
is also consistent with overproducing the mass of these galaxies at
$z=0$. A solution for this problem would be to assume a higher
mass-loading for low mass halos. For example, a recent study of
\citet{Dave13} has demonstrated that an energy-driven wind model for
dwarf galaxies (mass loading scaling with $\sigma^{-2}$) can reduce
the amount low-mass galaxies and thus, match the low-mass end of the 
stellar mass function better. Nevertheless to summarise, the overall
evolution of the peaks of the SFRs with halo mass (i.e. the stellar
downsizing) as predicted by the SHAM method is quantitatively very
well captured by the WM model.      

\begin{figure*}
\begin{center}
  \epsfig{file=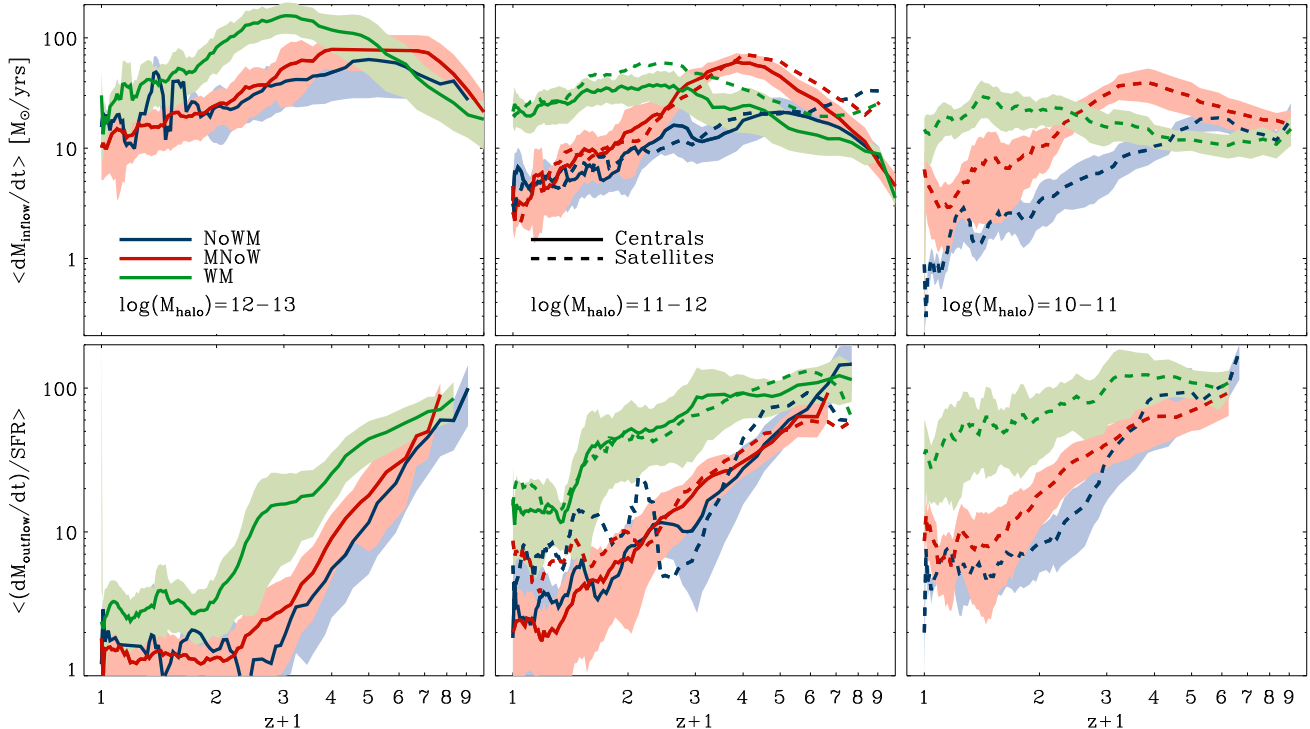,
    width=1.0\textwidth} 
  \caption{Cosmic evolution of the inflow rates (top row) and the
    mass loading factors (= $ \langle \dot{M}_{\mathrm{outflow}}/SFR
    \rangle$, bottom row) for different halo mass bins (different
    columns) and distinguishing between centrals and satellites (solid
    and dashed lines, respectively). Different colors illustrate the
    NoWM (blue), the MNoW (red) and the WM (green) simulation
    runs. The lines indicate the mean value, while the corresponding
    shaded areas show the 1-$\sigma$ scatter. In general, at $z
    \lesssim 3-4$, galactic winds increase the gas inflow (and outflow)
    rates and thus, lead to high mass loading factors.} 
 {\label{inout_evol}}
\end{center}
\end{figure*}

In the bottom row of Fig. \ref{SFR_evol}, we show the evolution of the
specific star formation rates $sSFR = SFR/M_{\mathrm{stellar}}$.
Overall, the sSFR's are continuously decreasing towards low redshifts.
The MW model shows the slowest decrease, in particular for galaxies in
lower mass halos. This is in good agreement with observations
presented by \citet{Dutton10, Elbaz11} and \citet{Reddy12} indicating
that sSFR's $\propto (z+1)^{2.25}$ below z = 2.5 and flat sSFR's at
higher redshifts.   

\begin{figure*}
\begin{center}
  \epsfig{file=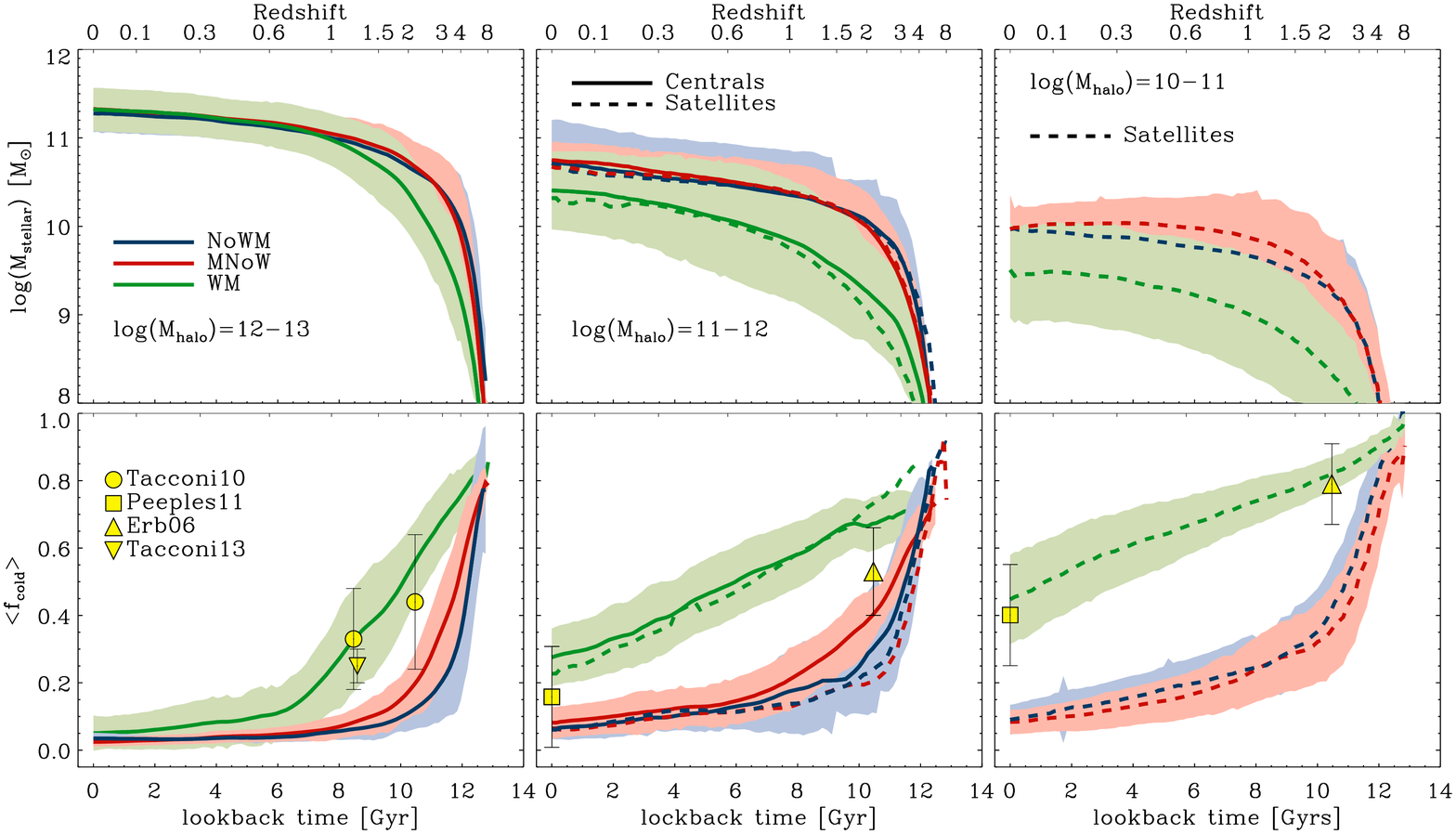, width=1.0\textwidth}
  \caption{Time evolution of the average stellar mass (top row)
   and the average cold gas fractions (bottom row) of the simulated
   galaxies in different halo mass bins (different columns) separated
   by centrals (solid lines) and satellites (dashed lines). Different
   colours correspond to the three NoWM (blue), MNoW (red) and
   WM(green) models. The shaded areas show the 1-$\sigma$ scatter of
   the mean values. Runs including momentum-driven winds lead to
   significantly higher cold gas fractions, particularly in low mass
   galaxies, which are in reasonably good agreement with observational
   data (yellow symbols, \citealp{Erb06, Tacconi10, Peeples11, Tacconi13}).}  
 {\label{fcold_evol}}
\end{center}
\end{figure*}

In Fig. \ref{sSFR_Mgal}  we show the sSFR's of the three different
models as a function of galaxy stellar mass at $z = 0, 1, 2$ (from top
to bottom) in comparison to observations. The solid and dashed  black
lines indicate observational results from different surveys (SDSS \& 
GOODS \& Galex: \citealp{Elbaz07, Daddi07, Salim07}) and the yellow
symbols illustrate the observational results from recent studies by
\citet{Tacconi10} and \citet{Tacconi13}. Here we have separated the
simulated galaxies into quiescent (light green, red and blue symbols) and
star-forming (dark green, red and blue symbols) using the \citet{Franx08}
definition: galaxies are quiescent for $\mathrm{sSFR} <
0.3/t_{\mathrm{Hubble}}$. While the NoWM and MNoW runs  in general
under-predict the sSFR's of star-forming galaxies at a given stellar
mass, the sSFR's in the WM simulations are in reasonably good agreement
with the observed trend although the absolute SFRs (at least for
massive systems) seem to be too high. This is in qualitative agreement
with \citet{Dave11a} who find that (for large cosmological boxes) the
WM model significantly over-estimates the number density of galaxies
at the massive end of  the stellar mass  function  ($M_{\mathrm{gal}}
\geq  10^{11}  M_\odot$) and also the high end of the SFR distribution
function. They  attribute this  failure to a missing  mechanism to
efficiently suppress  star  formation  in \textit{massive} galaxies
(e.g. feedback from AGN)  primarily at  low redshift.    
However, for lower mass galaxies ($M_{\mathrm{gal}} <5\times
10^{10} M_\odot$), we obtain slightly higher sSFR's at low redshifts
than the previous results of \citet{Dave11a}, a possible effect of the
higher resolution (i.e. increasing star formation with 
increasing resolution, see Appendix \ref{resolution}). 

\subsection{Inflow, outflow and mass loading of gas}\label{inoutflow} 

To explicitly demonstrate the influence of the galactic winds and
fountains on the balance of gas inflow and outflow we present in 
Fig. \ref{inout_evol} the cosmic evolution of the mean gas inflow
rates (top row), and of the 'mass-loading' factors (bottom row) for
star forming centrals and satellites in different halo mass bins. We
consider the flows through a sphere with radius $r_{10}$. The
'mass-loading' is then computed as 
\begin{equation}
\left\langle \frac{\dot{M}_{\mathrm{outflow}}}{SFR} \right\rangle,
\end{equation}
where $\dot{M}_{\mathrm{outflow}}$ is the outflow rate. Note that
the evolution of the outflow rates is not shown explicitly but their
behaviour follows the one of the inflow rates.

We find that at high redshifts ($\gtrsim 3-4$) the inflow (and
outflow) rates in the models without winds are higher than in the WM
model. This most likely reflects the inflows (and outflows) of hot gas
which appear in a hydrostatic equilibrium (of a hot halo) in the models
without winds. The early emergence of a hot halo is a consequence of
the earlier assembly of stellar structures in the NoWM and the MNoW
models than in the WM model and of the resulting gravitational energy
release from in-falling stellar systems (i.e. gravitational feedback,
see \citealp{Johansson09} and \citealp{Naab07}).   
In contrast, at redshifts below $\lesssim 3$, the inflow (and outflow)
rates in the WM model are always larger than in the models without
galactic winds. The high outflow rates are a direct consequence of
expelling the star-forming gas in the wind model, while the high
inflow rates reflect the accretion of recycled gas, which has been
blown out at earlier times and which is according to
\citet{Oppenheimer10} the dominant mode of gas accretion since $z=1.5$  
(besides cold and hot mode accretion). In the highest halo mass bin
(left top panel) the inflow (and outflow) rates in the WM model are
strongly declining below $z \sim 2$ and reaching at $z=0$ similar
values as in the NoWM and the MNoW models, whereas in galaxies
residing in halos with masses below $10^{11} M_\odot$, the inflow (and
outflow) rates remain almost constant below $z \sim 2$ and are at
$z=0$ by roughly an order of magnitude higher than in the models
without any galactic winds.  

Turning now to the mass loading factors (bottom row in Fig.
\ref{inout_evol}), all models predict a decreasing mass loading with
decreasing redshift. At very high redshifts $z>5$ the mass loading is
similarly high in all three simulation runs. With decreasing redshift,
however, the mass loading factors in the NoWM and MNoW models exhibit 
a similarly strong decline, while for the WM model the decline is much 
shallower and thus, the mass loading factors are always higher than in
the models without galactic winds, particularly for low-mass
halos. This explicitly demonstrates the efficiency of stellar
winds on the strength of mass loading (by definition). Only for the
most massive halos the mass loading factors in the  WM run at $z=0$
are found to be similarly low as those in the runs without winds as
the galactic wind feedback is less efficient in blowing gas out of
massive galaxies at low redshifts and as the SFRs are still high (due
to the high re-infall rates of previously ejected gas).

\subsection{Cold gas fractions}\label{cold} 

\begin{figure}
\begin{center}
  \epsfig{file=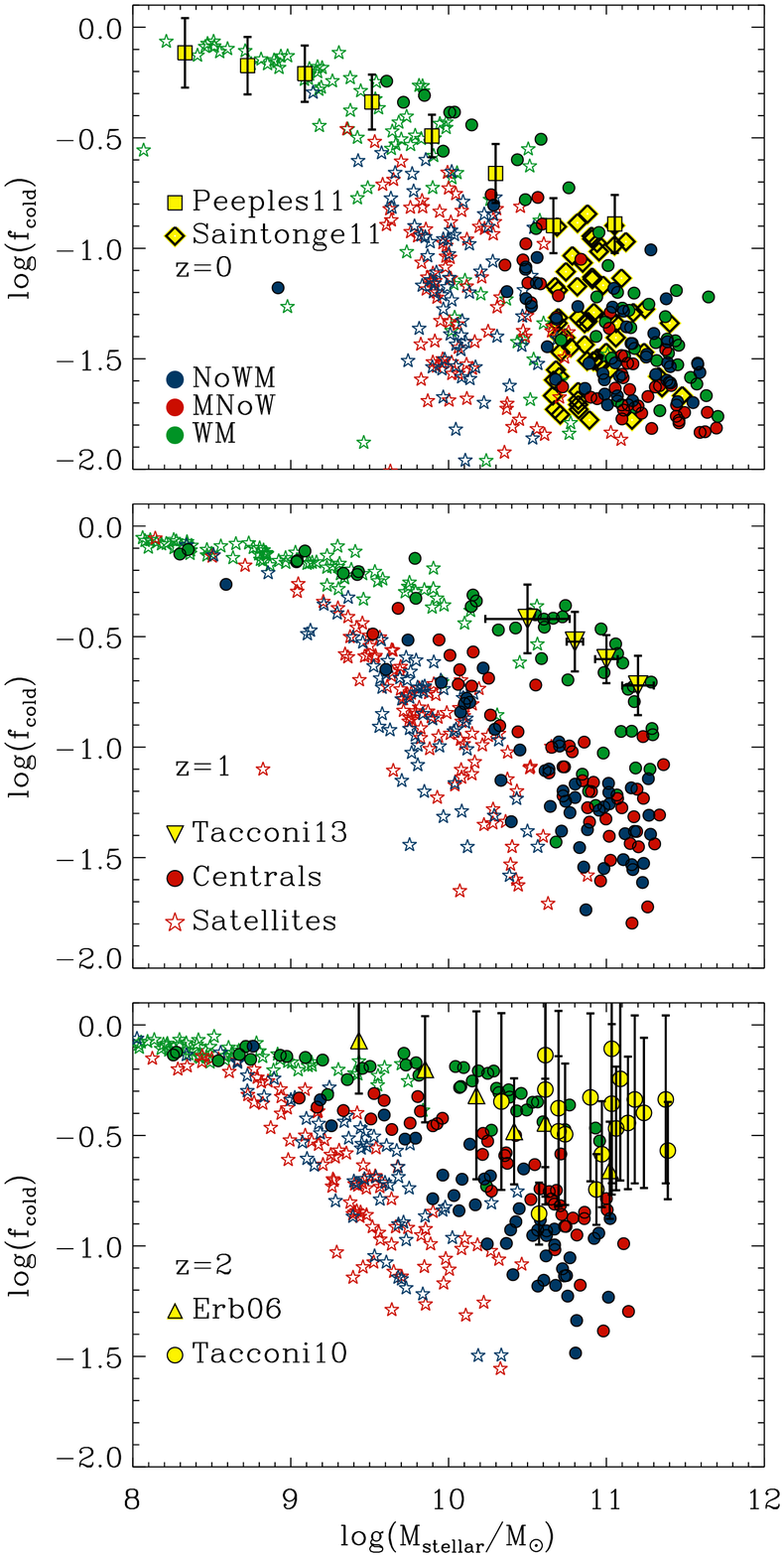, width=0.4\textwidth}
  \caption{Cold gas fractions versus galaxy masses at
    z=0, 1, 2 (top, middle and bottom panel) for the three
    models (indicated by different 
    colours). Filled circles indicate central galaxies, while open stars
    correspond to satellite galaxies. Yellow triangles, squares and
    circles show observations from \citet{Tacconi13, Peeples11,
      Tacconi10, Erb06}, respectively. The WM runs including metal
    cooling and winds can match the observational data at all
    redshifts fairly well.}  
 {\label{fcold_Mgal}}
\end{center}
\end{figure}

\begin{figure}
\begin{center}
  \epsfig{file=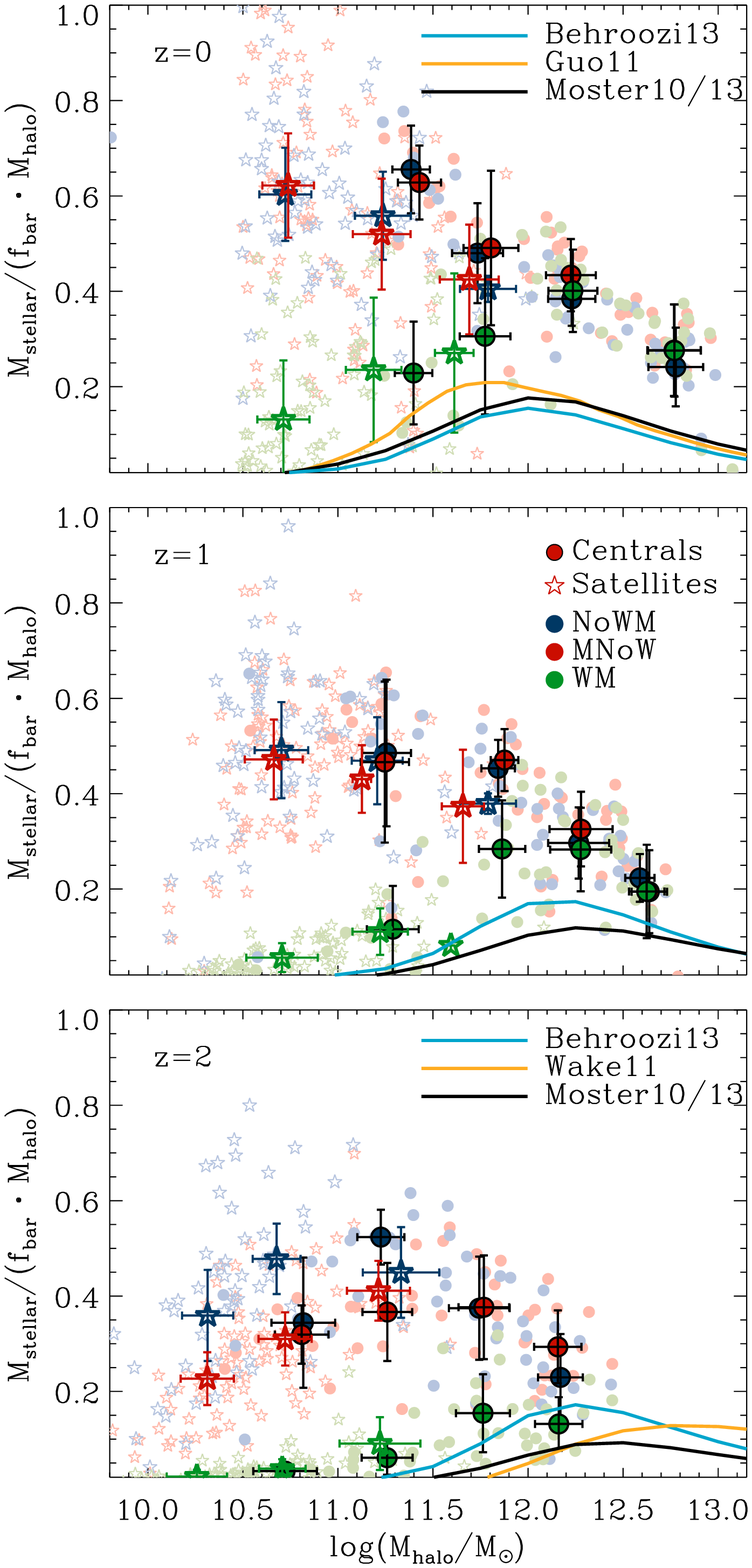,
    width=0.4\textwidth} 
  \caption{Baryon conversion efficiencies at z=0, 1, 2 (top, middle
    and bottom panel) for the three models (indicated by different
    colours). Filled circles and open stars indicate central and
    satellite galaxies, respectively. Solid lines show the predictions 
    from different abundance matching models of \citet{Moster10, Guo11,
      Wake11,Behroozi12,Moster12}. At z $\sim$ 2, the WM
    model matches the predicted conversion efficiencies fairly well at
    all halo masses. At z = 0, all simulation sets predict too high
    conversion efficiencies, in particular the models without winds
    (NoWM, MNoW) at low halo masses. The wind model WM accounts for
    the shape but overall predicts higher galaxy masses by a factor of
    2 to 4. } 
 {\label{Conveff}}
\end{center}
\end{figure}

In Fig. \ref{fcold_evol} we show the evolution of the stellar 
mass (top row) and the cold gas fractions (bottom row). The cold gas  
fraction is computed as     
\begin{equation}
f_{\mathrm{cold}} =
\frac{M_{\mathrm{cold}}}{M_{\mathrm{cold}}+M_{\mathrm{stellar}}},
\end{equation}
where $M_{\mathrm{cold}}$ is the cold gas mass and
$M_{\mathrm{stellar}}$ is the stellar mass within $r_{10}$. We find a
general trend for lower mass galaxies to have higher cold gas
fractions. For the NoWM and MNoW runs the evolution of the galaxies is 
very similar. The cold gas fractions at all masses decrease very
rapidly with time so that already by $z \approx 2$ the cold gas
fractions are below 20\%. Due to very efficient early star formation
most of the gas is consumed early-on. Including metal enrichment only
slightly increases the cold gas masses (and fractions) due to enhanced
cooling. It is notable that in both models the cold gas fractions of
the satellites decrease faster than for the centrals at the
\textit{same} halo mass. This may be caused by environmental processes
like strangulation, ram-pressure stripping etc. working only on   
satellite galaxies (e.g. \citealp{Gunn72, Abadi99, Quilis00,
  vanGorkom04, Roediger07, Kawata08, McCarthy08, Font08, Bekki09} and 
references therein). For the WM runs, the cold gas fraction are
significantly higher than for the NoWM and MNoW runs. 
In particular at $z \approx 2$ the winds result in 4-5 times higher
cold gas fractions. This is caused both by \textit{lower stellar}
masses (top row of Fig. \ref{fcold_evol}) and at the same time
\textit{higher cold gas} masses irrespectively of whether the galaxy
is a central or a satellite.  At higher redshifts ($z \gtrsim 3$), the
cold gas mass in the WM runs tends to be slightly lower than in the
models without winds (not explicitly shown), while the stellar masses
are significantly lower. Towards lower redshifts, however, we find
that the cold gas mass in the WM runs is \textit{always} larger than
in the NoWM and MNoW models. There is also a clear trend for lower
mass galaxies to have higher gas fractions. This is a consequence of
the re-accretion of recycled gas in the WM model
(\citealp{Oppenheimer10, Weinmann12}). The increased cold gas
fractions in the WM runs are caused by suppression of early star
formation, delaying the conversion of gas into stars towards lower
redshifts as galactic winds are very efficient in blowing out
star-forming gas particularly for low-mass galaxies which can be
re-accreted at later times and thus, leading to an increased amount of
cold gas compared to the models without galactic winds
(e.g. \citealp{Oppenheimer06, Stinson06, Dave11a,  Stinson12,
  Stinson13}. In general, the gas fractions of the MW model agree much
better with observations (here we have not distinguished between
molecular and atomic hydrogen). We have indicated respective
observations \citep{Tacconi10, Peeples11, Erb06}) by yellow symbols on
the bottom panels of Fig. \ref{fcold_evol}.  

A more consistent comparison of simulated and observed gas fractions 
and stellar masses at $z = 0 , 1, 2$ is presented in Fig.
\ref{fcold_Mgal}. Filled circles illustrate central, open stars
satellite galaxies in simulations. Yellow symbols correspond to 
different measurements of cold gas fractions as indicated in the
legend (\citealp{Erb06, Tacconi10, Saintonge11a} and
\citealp{Tacconi13}\footnote{We have used the incompleteness corrected 
  $z \sim 1$ data from the  \citealp{Tacconi13} PHIBSS sample.}) In
agreement with previous results \citep{Dave11b} the NoWM and MNoW 
runs consistently under-predict observed gas fractions while the
empirical wind models are in reasonable agreement with observations at
all redshifts and all galaxy masses.

\subsection{Baryon conversion efficiencies}\label{Confeff} 

Fig. \ref{Conveff} shows the baryon conversion efficiencies
versus dark matter halo mass at $z = 0, 1, 2$ compared g
to results from different abundance matching models (black, light 
blue and orange lines) of \citet{Moster10, Guo11, Wake11} and
\citet{Behroozi12,Moster12}. The baryon conversion efficiency
is defined as:  
\begin{equation}
\mathrm{Conversion}\ \mathrm{efficiency} =
\frac{M_{\mathrm{stellar}}}{f_{\mathrm{bar}} \cdot M_{\mathrm{halo}}},
\end{equation}
where $f_{\mathrm{bar}} \approx 0.17$ is the universal baryon
fraction. The lightly colored symbols in Fig. \ref{Conveff} illustrate 
the individual values for the simulated galaxies, while the full 
coloured, large symbols show the binned values with a 1-$\sigma$
scatter.  

At $z=0$, massive central galaxies are at least two to four times more
massive than predictions from abundance matching, independent of the
assumed model (see also \citealp{Oser10} and \citealp{Hirschmann12}).
The origin for the over-production of stars is, however, different. In
the NoWM and the MNoW models, too many stars are already formed at
high redshift, in particular in low mass halos (see Fig. \ref{SFR_evol} 
and middle and bottom panel of Fig. \ref{Conveff}). In the WM model,
there is too much in situ star formation at redshifts $z \leq 1$
caused by the infall of recycled gas and possibly missing AGN feedback
(see \citealp{Dave09}). Increasing the mass loading ($\sigma_0 \approx
600-900\ \mathrm{km/s}$) or the kick velocity does slightly lower the
present day baryon conversion efficiencies but reduces the SFR at
$z>1$ too much. This indicates that the WM model cannot fully account
for processes driving the evolution of galaxies in massive halos
($M_{\mathrm{halo}} > 10^{12} M_\odot$). There is only a small
difference in the baryon conversion efficiencies of the NoWM and the
MNoW models, in contrast to semi-analytic models which normally
predict larger differences (\citealp{Hirschmann12}). Towards lower
masses only the MW runs follow the predicted relation with a
comparable constant offset. Using a similar wind model, \citet{Dave09}
find lower baryon conversion efficiencies for present-day, low-mass
halos. This can be interpreted as a consequence of their lower spatial
resolution (see Appendix \ref{resolution}). Both models without winds
significantly over-predict the stellar masses at higher redshifts. At
$z \approx 2$ the galaxy stellar masses of the wind model agree with
the abundance matching constraints very well both at high and low
masses and for satellites and centrals due to the efficient
suppression of early star formation (see Fig. \ref{SFR_evol}). The
difficulty of predicting low baryon conversion efficiencies in dwarf
galaxies and matching observed galaxy abundances was already raised in
a previous study by \citet{Sawala11} who stated that dwarf galaxies
formed in their and all other hydrodynamical simulations at that time
are more than an order of magnitude more luminous than expected for
haloes of this mass. The importance and the success of adopting a
strong stellar feedback in simulations to overcome this problem and to
regulate the star formation efficiencies over a large range was also
demonstrated by other recent studies such as the one of
\citet{Munshi13} using the \textsc{Gasoline} code with a blastwave
supernova feedback.

\begin{figure*}
\begin{center}
  \epsfig{file=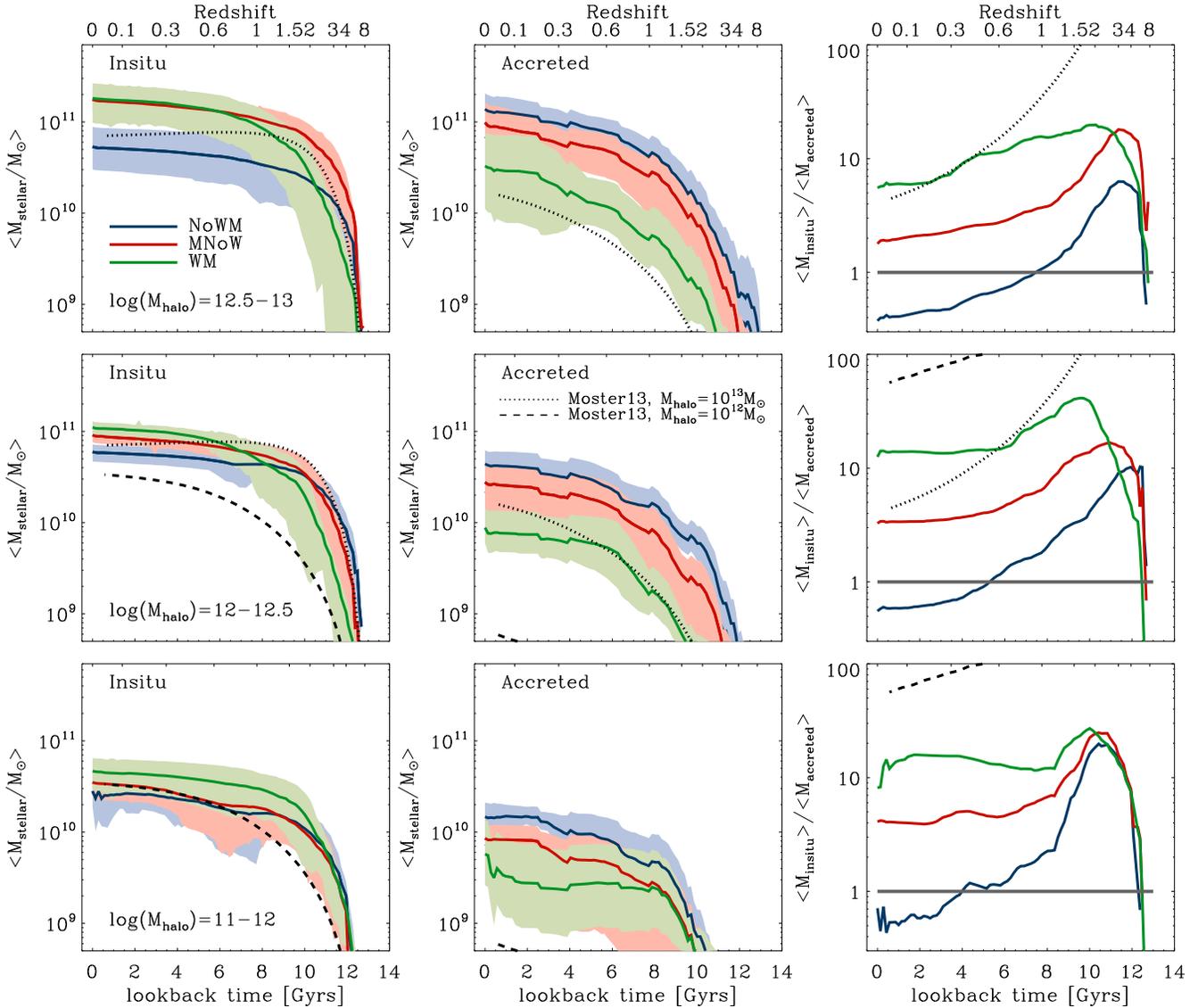, width=1.0\textwidth}
  \caption{Time evolution of in situ formed stars (left column), accreted
    stars (middle column) and the fraction of the two (right column)
    in three different halo mass bins (from top to bottom, considering
    the \textit{present-day} halo masses). Different
    models are illustrated by different colours. The solid lines show
    the average value, while the shaded areas illustrate the
    1-$\sigma$-scatter. Both metal cooling (MNoW) and galactic winds
    (WM) increase the in situ/accreted ratio due to a smaller amount
    of accreted stars and a higher in situ star formation than in the
    NoWM runs. Model results are compared to the predictions from the
    SHAM approach (\citealp{Moster12}, black dashed and dotted lines
    for $10^{12} M_\odot$- and $10^{13} M_\odot$-mass halos,
    respectively).} 
 {\label{MaccIns}}
\end{center}
\end{figure*}

\section{The two modes of stellar mass assembly}\label{AccIns} 

Galaxies can grow their stellar masses in two ways: by converting cold
gas into stars in situ in the galaxy or by accreting stars that have
formed in other galaxies in mergers \citep{Abadi03, Naab07,
  Johansson09, Oser10}. We refer to these two modes as 'in situ' and
'accreted'. The relative amount of in situ and accreted stars was found
to vary systematically with galaxy mass in simulations, for
semi-analytical models as well as for estimates from abundance
matching \citep{Oser10, Lackner12, Gabor12, Johansson12,
  DeLucia06, Guo08, Hirschmann12, Moster12, Behroozi12, Yang13}.     

In Fig. \ref{MaccIns} we show the time evolution of the mass of
in situ formed stars (left panels), accreted stars (middle panels) and
the ratio of the two (right panels) for three halo mass bins. Note that 
we have only considered central galaxies here and the halo mass bins
have changed with respect to previous figures. The abundance matching
results from \citet{Moster12} are indicated by the black lines.  The
NoWM galaxies grow rapidly and in situ star formation dominates at
high redshift $z>1$ followed by accretion of stars at $z<1$. The
contribution from accreted stars can dominate the overall mass budget
of the most massive galaxies. Already the inclusion of metal cooling
reduces the importance of accretion: high-mass galaxies grow faster at
high redshifts due to enhanced cooling and the accreted stellar mass
is slightly reduced (also because the accreted galaxies are slightly
more dominated by gas, see Figs. \ref{fcold_evol} and
\ref{fcold_Mgal}). However, galactic winds have the most dramatic
effect. For all galaxies the amount  of accreted stars is
significantly reduced as the merging halos host galaxies with
significantly higher gas masses and lower stellar masses (see
Figs. \ref{fcold_evol} and \ref{Conveff}). As a result, galaxies in
halos at $\sim 10^{12} M_{\odot}$ accrete less than 10  percent of
their stars. The ratio of in situ formed stars is still high for
massive galaxies. Here the galaxies are, however, influenced by the
potentially unrealistically high star formation rates. These may be
reduced by additional feedback from AGN, compensating the effect of
stellar winds and metal cooling (e.g.  \citealp{Croton06, 
  Bower06,Cattaneo06, Kang06, Monaco07,
  Bower08,Somerville08}). \citet{Hirschmann12} demonstrated that the 
fraction of in situ formed to accreted stellar mass is reduced by a
factor of $\sim 4$ in massive galaxies when AGN feedback is
considered using semi-analytic models. Recently, \citet{Dubois13} also
indicated with direct simulations that AGN feedback can suppress late
in situ star formation leading to a dominance of accretion of stars in
massive galaxies with halo masses $4 \times 10^{12} <
M_{\mathrm{halo}} < 8 \times 10^{13} M_\odot$ (with similar accreted
fractions as in \citealp{Oser10}). This clearly suggests that feedback
from AGN and metal cooling/galactic wind feedback in massive galaxies 
seem to have \textit{compensating} effects on fraction of accreted
stars.  

\begin{figure}
\begin{center}
  \epsfig{file=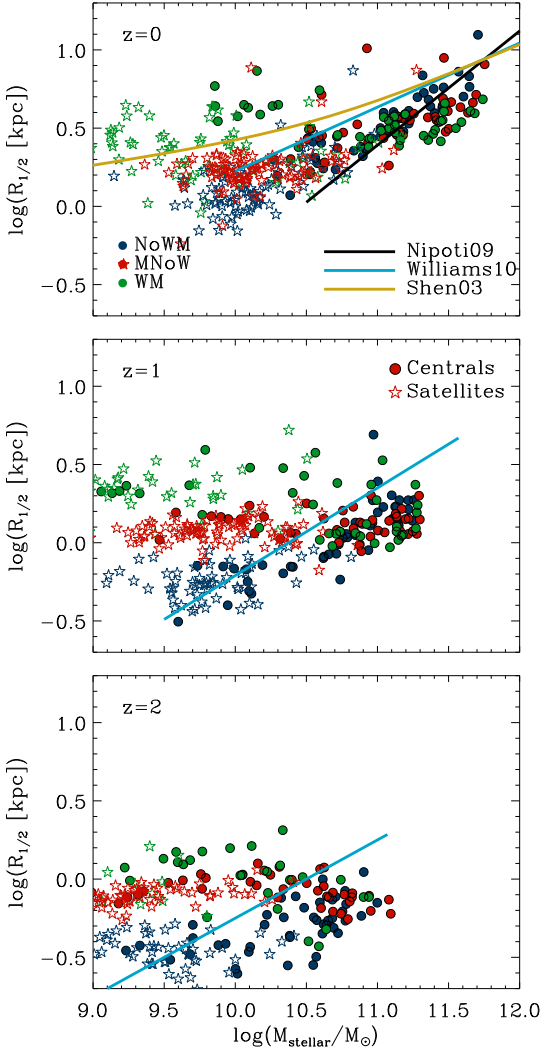, width=0.4\textwidth}
  \caption{Mass-size relation of central (circles) and satellite
    galaxies (stars) at $z= 0,1,2$ (top, middle and bottom panel) for
    the three models (indicated by different colours). The half-mass
    radii have been computed using the projected stellar half-mass
    radius within $r_{10}$. Simulation results are compared to two
    different observed relations for \textit{elliptical} galaxies
    (\citealp{Nipoti09, Williams10}) and \textit{spiral} galaxies
    (\citealp{Shen03}). High-mass galaxies are typically smaller using
    the WM model which over-predicts late in situ star formation,
    while low-mass galaxies are typically disk-dominated and thus,
    larger in the WM than in the NoWM model.} 
{\label{Mass_size}}
\end{center}
\end{figure}

\begin{figure}
\begin{center}
  \epsfig{file=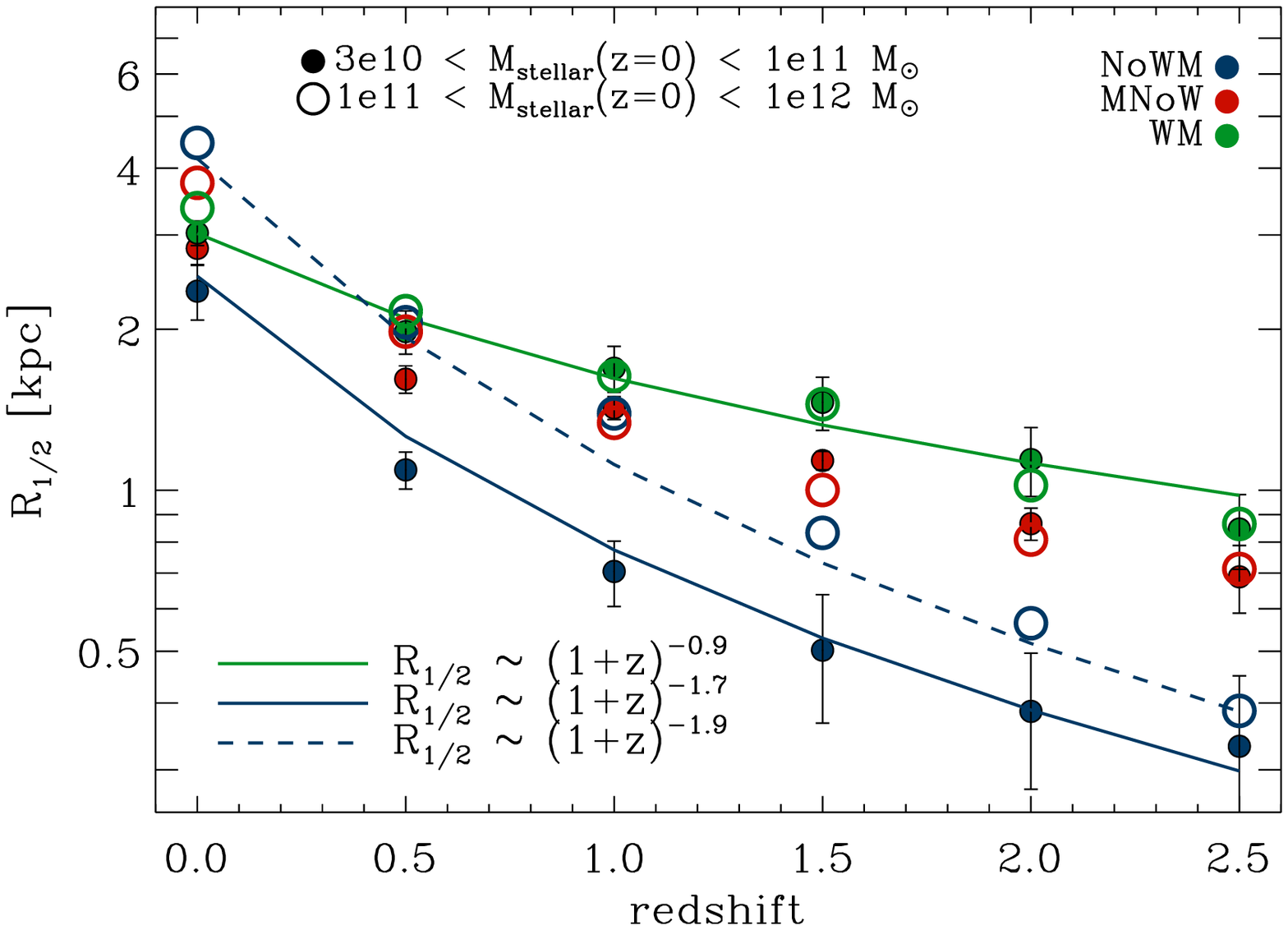, width=0.45\textwidth}
  \epsfig{file=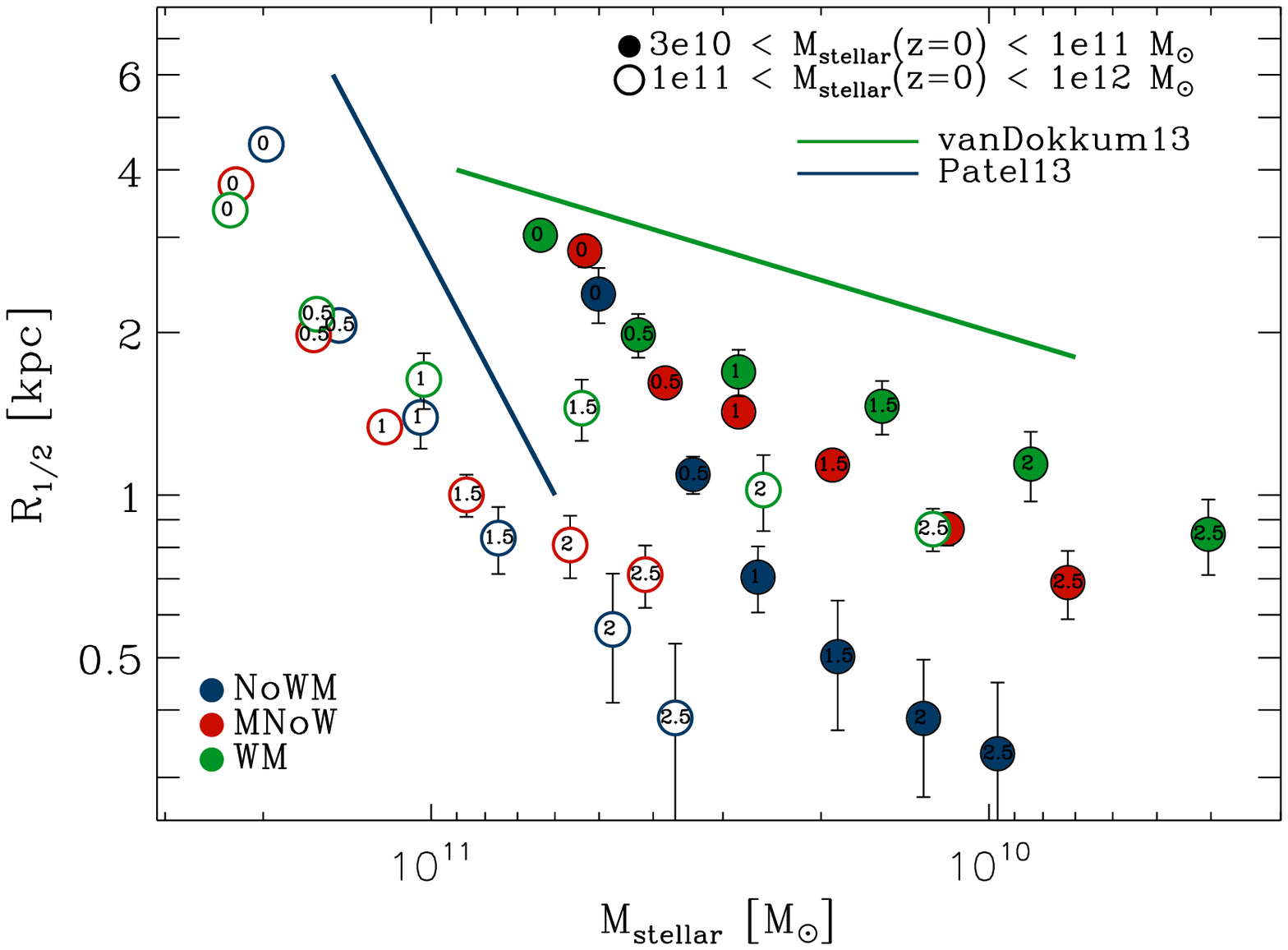, width=0.45\textwidth}
  \caption{Mean half-mass radii versus redshift (top panel) and
    versus stellar mass of the progenitor galaxies (bottom panel)
    considering only central galaxies binned according to their
    present-day stellar mass (filled circles: $3 \times 10^{10} <
    M_{\mathrm{stellar}} < 10^{11} M_\odot$, open circles: $10^{11} <
    M_{\mathrm{stellar}} < 10^{12} M_\odot$). Different simulation
    models are shown by different colours. The green and
    blue, solid and dashed lines in the top panel show a fit to the WM
    and NoWM models for the two different mass bins. The numbers
    within the circles in the bottom panel indicate the redshift. The
    size evolution for one simulation model hardly changes when
    varying the present-day stellar mass with the consequence that for
    the massive galaxies, only the NoWM model can capture the strong,
    observed size evolution (blue solid line in bottom panel,
    \citealp{Patel13}), while the weaker size evolution in the WM
    model is in rough agreement with the observed slope of
    intermediate mass galaxies (green solid line in bottom panel,
    \citealp{vanDokkum13}).    
  }  
{\label{Size_evol}}
\end{center}
\end{figure}

Turning now to the in situ formed fraction of stellar mass, the
studies by \citet{Moster12} and also by \citet{Yang13} (using the
conditional stellar mass function, the CSMF model based on local
observational constraints of the star formation rates of central
galaxies) find for halo masses of $M_{\mathrm{halo}} \sim 10^{13}
M_\odot$ present-day, in situ fractions of 80 and 75~per  cent,
respectively, which is in good agreement with our value  of 85 per
cent for the WM model (see black dotted and green solid lines in the
top right panel of Fig. \ref{MaccIns}). Instead, our in situ fractions
are somewhat too high compared to the results of \citet{Behroozi12}
who apply a similar method as \citet{Moster12}, but obtain lower in
situ fractions between 30 and 40~per cent. At halo masses below 
$M_{\mathrm{halo}} < 3 \times 10^{12} M_\odot$, \citet{Moster12} and
\citet{Yang13} also predict higher, present-day in situ fractions
between 85 and 99 per cent, in qualitative agreement
with the $\sim93$ per cent in the WM run, while the results of
\citet{Behroozi12} predict lower in situ fractions of 55 and
80~per cent for $10^{12}$- and $10^{11} M_\odot$-mass galaxies,
respectively. However, for the lowest halo mass bin the in situ
fractions in the WM model are low compared to the SHAM prediction (see
green solid and black dashed  lines in the bottom right panel of Fig.
\ref{MaccIns}). Overall, this analysis demonstrates that metal cooling
and strong winds seem to be necessary and essential to produce high in
situ fractions (as derived from observations) in lower mass halos. The
significant deviations to the SHAM/CSMF model - which do reproduce the
cosmic star formation rate history - also indicate the need for an
improved treatment of galactic winds (i.e. stronger suppression of
early star formation in low-mass halos). A possible, purely
observational test for the accreted versus in situ formed stellar
fraction are abundance gradients at large radii ($r > r_{\mathrm{eff}}$).  
Here, the models predict that accreted stars assemble typically at
larger radii and originate from lower mass galaxies with lower
metallicities. Therefore steep stellar metal gradients - at radii were
stars typically do not form - might indicate high accreted
fractions. Additional insight might come from the distribution of
globular cluster systems (see e.g. \citealp{Forbes11, Forbes12}). Only
the wind model  predicts such steep gradients which we will
demonstrated in a follow up study. 
\begin{figure}
\begin{center}
  \epsfig{file=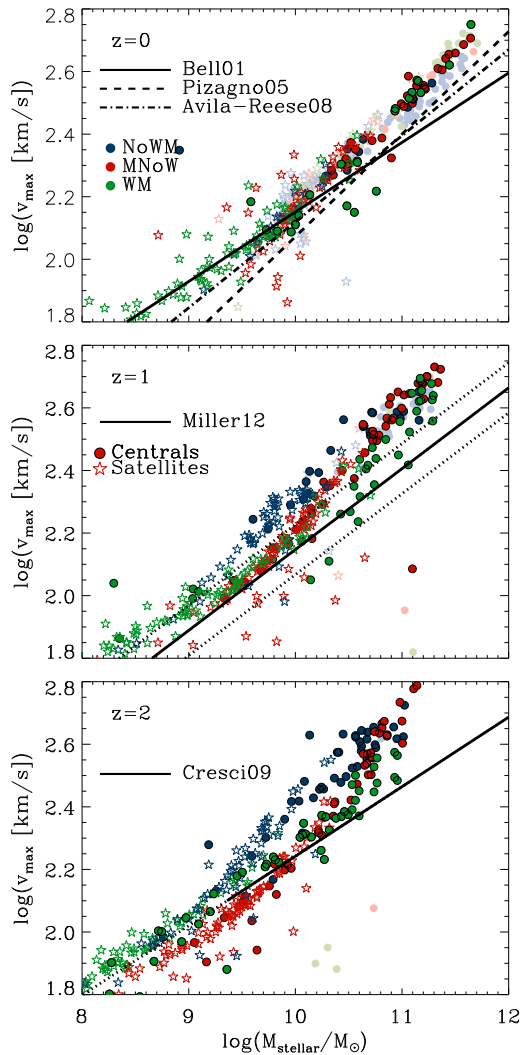, width=0.4\textwidth}
  \caption{Tully-Fisher relation at z = 0, 1, 2 (from top to bottom)
    for the three models distinguishing between star-forming (full
    colours) and quiescent galaxies (light colours). Filled 
    circles and open stars indicate central and satellite galaxies,
    respectively. The maximum rotation velocities have been computed
    at $r_{10}$ using the contribution from stars, gas and the dark
    matter halo. Simulation results are compared to observed relations
    from \citet{Bell01,Miller12} and \citet{Cresci09}.} 
{\label{TF}}
\end{center}
\end{figure}

Recently, \citet{Lackner12}, who employ an AMR code (ENZO) with metal
cooling  and thermal supernova feedback (and thus, corresponding to
our red curves), predicted a dominance of in situ star formation in
massive galaxies for the entire redshift range, in agreement with our
MNoW model, but in contrast to  \citet{Oser10} (whose accreted
fractions are twice as high). \citet{Lackner12} attribute these
discrepancies to the different modelling of supernova feedback which is
more efficient in their simulations. Numerical resolution and assumed
star formation efficiencies might play a significant role as well.

\section{Scaling relations}\label{Size} 
\begin{figure*}
\begin{center}
  \epsfig{file=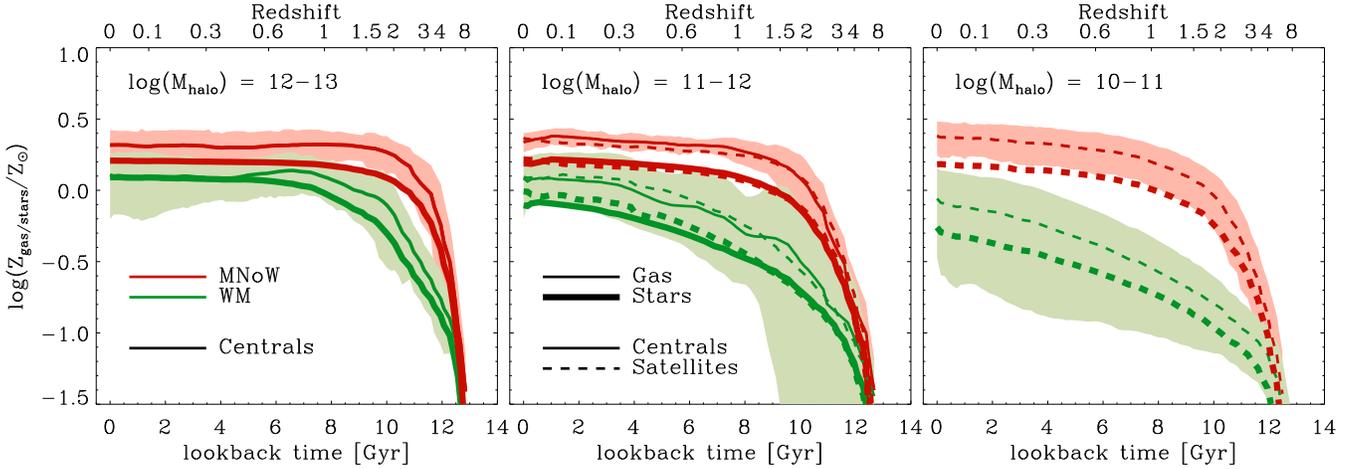, width=1.0\textwidth}
  \caption{Time evolution of the mean gas (thin lines) and stellar
    metallicity (thick lines) in different halo mass bins (different
    panels) and distinguishing between centrals (solid lines) and
    satellites (dashed lines). Different colours correspond to the MNoW
    (red) and WM (green) model. The shaded areas show the
    1-$\sigma$ scatter of the mean gas metallicity, the 1-$\sigma$
    scatter of the mean stellar metallicity is not shown explicitly,
    but is similar to the gas. In the WM runs gas and stars in the
    galaxies are less metal-enriched than in the MNoW runs and
    additionally, momentum-driven winds delay the metal enrichment
    of both the gaseous and the stellar phase, this effect increases
    with decreasing halo mass.}   
 {\label{GasMet_evol}}
\end{center}
\end{figure*}

Fig. \ref{Mass_size} shows the projected stellar half-mass radius
(i.e. the size) versus  galaxy mass for central (circles) and
satellite (stars) galaxies for the three models at $z=0,1$ and $2$
(different panels). The half-mass  radius of a galaxy is defined as
the \textit{projected radius, within  which half of the final galaxy
  mass within $r_{10}$  is  included}.

Compared to observations of early-type galaxies by
\citet{Nipoti09} or \citet{Williams10} (black and blue solid lines),
the NoMW model predicts reasonable sizes for massive galaxies
($M_{\mathrm{stellar}} > 10^{11} M_\odot$, top panel of Fig.
\ref{Mass_size}), in agreement with \citet{Oser12}. In contrast,
massive galaxies in the simulation runs including metal cooling and/or 
winds (where a larger fraction of the galaxies is star-forming) have
sizes which are about $\sim 0.2$ dex too small compared to the
observed most massive galaxies $\log(M_{\mathrm{gal}}/M_\odot) >
11.3$.  This is most likely a direct consequence of the reduced
fraction of accreted stellar mass and the high in situ star formation
at all times (see section \ref{AccIns}), which significantly
suppresses the size growth at low redshifts. This result is in
agreement with a recent study of \citet{Dubois13} who show that
simulations without AGN feedback produce too small sizes as a
consequence of a high, late in situ star formation. In their work,
they attribute AGN feedback to be responsible for a significant growth
of the size of a massive galaxy since $z \sim 2$ as a direct result of
a significantly increased accretion of stars at late times and a
reduction of star formation at the galaxy centre due to the released
energy by the AGN. As for the two modes of the stellar mass assembly,
this demonstrates clearly (and maybe not surprisingly) the opposite
effects of AGN feedback and metal cooling/stellar winds on the size
evolution of massive galaxies. Overall, even if the NoWM model can 
successfully reproduce the mass-size relation as 
opposed to the WM  model, it does not ``justify'' the NoWM model but
instead indicates a missing process in the WM model which is most
likely feedback from AGN.  However, the most massive galaxies at
higher redshifts ($z=1,2$ in the middle and bottom panel of
Fig. \ref{Mass_size}), their half-mass radii at a given stellar mass,
do not vary much for the different models and even compared to
observations from \citet{Williams10}, all of them tend to be very
compact.  

At present, simulated galaxies in the NoWM and MNoW model with stellar 
masses of $M_{\mathrm{stellar}} < 5 \times 10^{10} M_\odot$ (top panel in
Fig. \ref{Mass_size}) are small compared to the observed mass-size    
relation for late-type galaxies (beige line, \citealp{Shen03}).
Instead, galaxies simulated with the WM model have a factor of two to
three larger sizes and agree well with the observed sizes. Also at
higher redshifts ($z=1,2$ middle and bottom panel of
Fig. \ref{Mass_size}), the WM model results in larger low mass
galaxies resulting in an interesting turnover at redshift z =2: low
mass galaxies are larger than high-mass galaxies. In a study of
\citet{Brooks11} using high-resolved disk simulations performed with
\textsc{Gasoline} (including a blastwave feedback scheme), they can
also demonstrate that their simulated disks provide an excellent match
to the observed magnitude-size relation for both local disks and for
disks at $z = 1$. 

The increased sizes in the MNoW model (compared to the NoMW model) are   
most likely a consequence of efficient cooling so that high angular
momentum  gas in the outer parts of a galaxy can already form 
stars. The even larger sizes in the WM model indicate  that a strong
wind feedback helps to transform smaller bulge-dominated into larger
disk-dominated galaxies preferentially removing low angular momentum
gas. The importance of strong winds in removing low angular momentum
gas was first described by \citet{Binney01} and later discussed
e.g. in \citet{Brook11}.

The top panel of Fig. \ref{Size_evol} shows the mean half-mass radii
versus redshift for massive central galaxies at $z=0$ with
$M_{\mathrm{stellar}} > 10^{11} M_\odot$ and their progenitors (filled
circles) as well as  for lower mass centrals with  $3 \times 10^{10} <
M_{\mathrm{stellar}} < 10^{11} M_\odot$ (open circles). For the NoMW
model, the size evolution is similarly strong for high and low mass
galaxies with $R_{1/2} \propto (1+z)^\alpha$ with $\alpha \sim  -1.8$
(see also \citealp{Oser12}) and the high mass galaxies are about a
factor $\sim$ 2 larger by $z =0$. In contrast, the size evolution of
the WM galaxies is much weaker ($R_{1/2} \propto (1+z)^\alpha$ with
$\alpha = -0.9$) and there is almost no size difference between the
two mass bins (see Fig. \ref{Mass_size}). The similarity to the MNoW
model indicates that more efficient cooling (due to metals) without
any additional galactic winds already significantly reduces the
cosmological size evolution.    

\begin{figure}
\begin{center}
  \epsfig{file=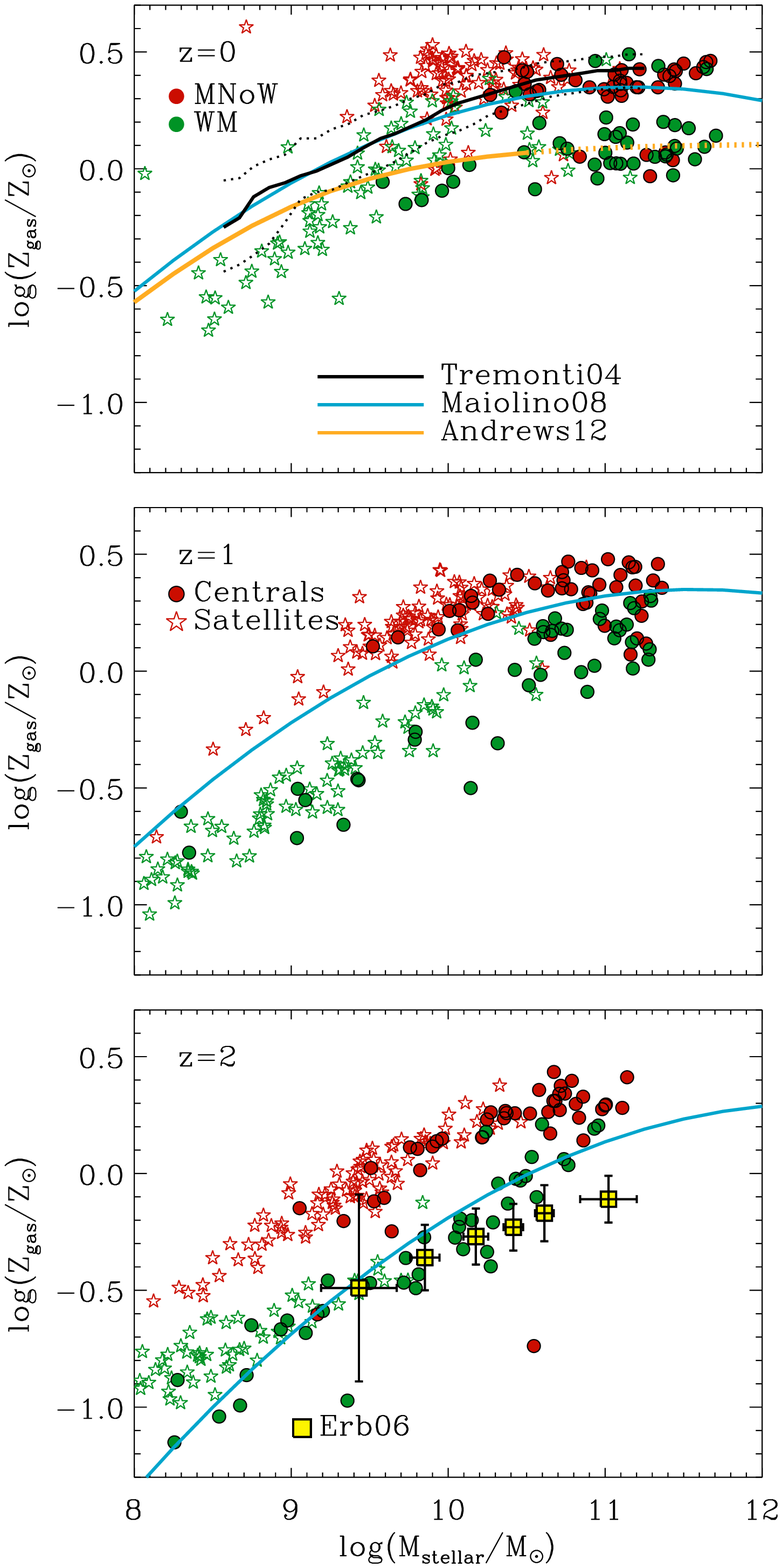, width=0.4\textwidth}
 \caption{Gas-phase metallicity versus galaxy mass at $z = 0, 1, 2$
   (from top to bottom) for the two models including metal enrichment
   (red, green). Filled circles and open stars indicate central and
   satellite galaxies, respectively. Yellow squares and solid lines illustrate
   measurements from \citet{Andrews12, Maiolino08, Erb06,
     Tremonti04}. WM runs reproduce the observational data best at
   $z=2$ and the observational data of \citet{Andrews12} at $z=0$, while
   the MNoW runs are in good agreement with observations of
   \citet{Tremonti04, Maiolino08} at $z=0$.} 
 {\label{GasMet_mass}}
\end{center}
\end{figure}
\begin{figure}
\begin{center}
  \epsfig{file=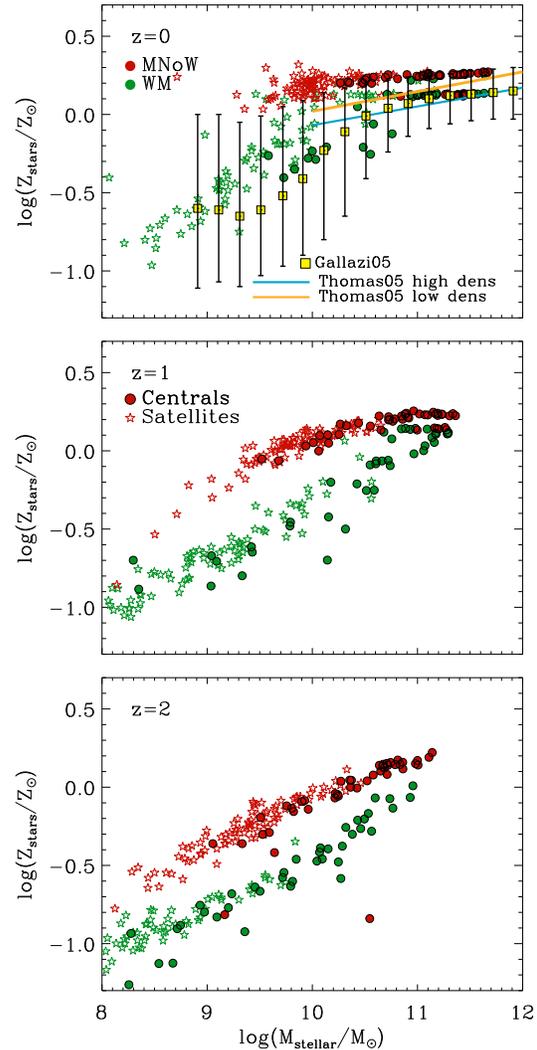, width=0.4\textwidth}
  \caption{Same as Fig. \ref{GasMet_mass}, but for stellar
    metallicity. Yellow squares and solid lines show observational
    data from  \citet{Thomas05, Gallazzi05} at z=0. WM runs are in
    good agreement with the observations. In particular the WM model
    reproduces the rapid decline in metallicity at masses $ < 10^{11} M_{\odot}$.} 
 {\label{StarsMet_mass}}
\end{center}
\end{figure}

Inspired by \citet{vanDokkum13} we show the galaxy size as
a function of stellar mass for high-mass (open circles) and low-mass
(filled circles) progenitors at different redshifts in the bottom
panel of  Fig. \ref{Size_evol}. Again, NoMW galaxies evolve rapidly 
$R_{1/2} \sim M_{\mathrm{stellar}}^2$ similar to observed massive
galaxies (blue line) but much steeper than observed lower mass
Milky-Way progenitors (green line) (\citealp{vanDokkum13,Patel13}). 
In contrast, the WM model again reveals a weaker size growth as a
function of mass with $R_{1/2} \sim M_{\mathrm{stellar}}^{0.4}$ almost
independent of mass and being in rough agreement with the observational  
trend. Compared to the recent observations of \citet{vanDokkum13} and 
\citet{Patel13} (dotted green and blue lines), we find that NoMW model
provides a good fit to the evolution of only massive galaxies whereas
the MW galaxies fit observed low mass galaxies. This is demonstrates
one of our major results in this work: none of the models can
reproduce the size evolution for both mass ranges at the same time.
The failure of the WM model to predict a steep size evolution of
massive galaxies is most likely due to their reduced fraction of
accreted stellar mass and their high in situ fractions at all times
and we may speculate that an additional AGN feedback may resolve this
mismatch.

Fig. \ref{TF} shows the relation between the circular velocity
of a galaxy and its stellar mass (Tully-Fisher relation,
\citealp{Tully77}) for the different models at  $z = 0, 1,
2$. Galaxies are separated into star-forming (full colors) and
quiescent (light colors) galaxies. We assume the  maximum circular
velocity $v_{\mathrm{max}}$ within $r_{10} $ and compare to observed
relations  from \citet{Bell01, Pizagno05,Avila-Reese08, Cresci09,
  Miller12}. The NoMW galaxies lie above the TF relation at all
redshifts due to the overproduction of centrally concentrated stellar
systems. MW galaxies, however, fit the observed TF at $z=2$ and, apart
from a turn-up  at $\log M_* > 10.5$, at $z=1$ and $z=0$ pretty
well. As for this model the stellar mass within a given halo is
reduced (see Fig. \ref{Conveff}), the baryons do not dominate the
circular velocity profile any more. For more massive systems the
unrealistically high late star formation moves that galaxies away from
the relation towards $z=0$. Our result is agreement with a previous
study of \citet{McCarthy12} who - using the GIMIC simulations with a
\textit{constant} wind model - can also reproduce the present-day
Tully-Fisher at stellar masses below $\lesssim 3 \times 10^{10}
M_\odot$ though, but obtain a similar 'kink' at higher stellar mass,
i.e. too large maximum circular velocities at a given stellar mass.
This demonstrates that when the simulated Tully-Fisher  relation is in
agreement with the observations (high redshifts and lower mass
galaxies at lower redshifts), the simulations also match the
predictions from the abundance matching models. At low redshifts and
for massive galaxies, the models are not in agreement with both the
observed TF relation and the abundance matching models anymore
(e.g. baryon conversion efficiencies). This result might contribute to
the reliability of the abundance matching models.

\section{Ages and metallicity}\label{Metal} 

Finally, we turn to the stellar and gas metallicity content of the
simulated galaxies. Fig. \ref{GasMet_evol} shows the time
evolution of the mean metallicity in the gas weighted by the SFR
(thin lines) and the stellar component weighted by stellar mass
(thick lines) within $r_{10}$, respectively, again binned into three
different halo mass bins and distinguishing between centrals and
satellites (solid and dashed lines, respectively). The
1-$\sigma$-scatter is indicated by the shaded areas, the one of the
stellar metallicities is not explicitly shown, but is similar to the
one of the gas metallicity.  Overall, the gas metallicities are
slightly higher than the stellar metallicities at a given time and
halo mass as the gas gets continuously  enriched. The WM runs predict
more metal poor (in the gaseous and stellar phase) galaxies than the
MNoW runs for the entire redshift and mass range. This is a natural
consequence of the galactic winds blowing metal-enriched gas out.  As
an indirect consequence of the blown-out metal-rich gas, the stellar
metallicity will be also reduced, since stars form out of the gas
present in the centre of a galaxy (\citealp{Scannapieco06, Wiersma09,
  Wiersma11, Tissera12, Tissera13}).

The observed and simulated stellar and gas metallicities as a function
of galaxy mass are compared in Figs. \ref{GasMet_mass} and
\ref{StarsMet_mass}. Black and coloured lines and the yellow
symbols illustrate different observational data from
\citet{Tremonti04, Erb06, Maiolino08, Andrews12, Thomas05} and
\citet{Gallazzi05}. Both simulations sets can reproduce a tight
correlation of both the stellar and gas-phase metallicities with
galaxy mass at all redshifts. The gas and stellar metallicities in the
MNoW runs are found to be always larger than the ones predicted by the
WM runs for the same reason as discussed above.  At $z = 2$, the gas
metallicities in the WM runs match the observational data well, while
the MNoW runs over-predict the gas metallicity for a given galaxy
mass. However, at $z = 0$ the situation is not clear as observations
can differ significantly depending on the calibration method (direct
mass-metallicity approach in \citealp{Andrews12} versus the strong
line calibration mass-metallicity in \citealp{Tremonti04}). The gas
metallicities in the MNoW runs provide a fair match to the
observational data of \citet{Tremonti04} and \citet{Maiolino08}, while
the ones from WM runs agree better with the overall lower gas-phase
metallicities from the study of \citet{Andrews12}. The lower gas
metallicities in the WM than in the MNoW runs may be caused by mixing
of metal-enriched gas with re-infalling metal-poor gas (in the WM
runs) which was blown out of the galaxy at earlier times. The stellar
metallicities predicted by the WM runs agree very well with the
observational data, in particular with respect to the observed decline
in metallicity towards the low mass end.  
We want to point out that the momentum-driven wind model and the
corresponding scaling of the mass loading with the velocity dispersion
is \textit{essential} for the reproduction of the steep decline in
both the stellar and gas metallicity towards the low stellar mass
end. \citet{Finlator08} and \citet{Dave11b} have demonstrated that
adopting a constant wind model, i.e. with a fixed mass loading and
wind velocity, produces an almost flat mass-metallicity relation
with a large scatter and is, thus, in strong disagreement with
observations.  

\begin{figure}
\begin{center}
  \epsfig{file=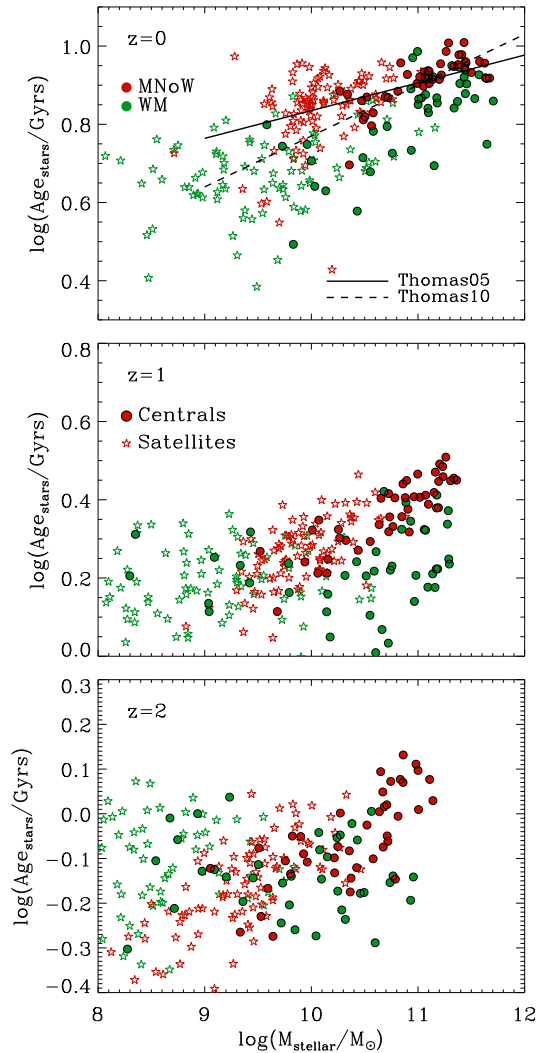, width=0.4\textwidth}
  \caption{Average mass-weighted stellar ages fo the simulated central
    (circles) and satellite (crosses) galaxies as a function of stellar
    mass. Black lines indicate observed relations from \citet{Thomas05,
      Thomas10}. The WM runs predict slightly younger galaxies than the
    MNoW runs, but both sets are in fairly good agreement with the
    observational data. At the high mass end the WM galaxies are
    consistently younger than observed because of the still on-going
    star formation at low redshifts.} 
{\label{Age_mass}}
\end{center}
\end{figure}

We compare the mass-weighted ages to observational data in
Fig. \ref{Age_mass} for $z=0,1,2$ (different panels). The MNoW runs
produce a positive correlation between stellar age and galaxy mass
(more massive galaxies are older) at all redshifts, but with a large
scatter in the stellar age. In contrast, at $z=1$ and $z=2$, the WM
runs produce stellar populations with ages being nearly independent of
galaxy mass and also having a large scatter. At the high mass end, the
stellar populations from the WM runs tend to be younger, while at the
low mass end they seem to be older than the ones from the MNoW
runs. The latter point is most likely due to the strong stellar
feedback which prevents star formation in low-mass galaxies so that
they are older. At present, the stellar ages predicted by the
WM runs are also related to the galaxy mass (with a large
scatter in the ages). But compared to the MNoW runs, the stellar
populations of the WM runs tend to be slightly younger what
is most likely a consequence increased late in situ star formation in
massive galaxies compared to the MNoW runs. However, both simulation
sets produce stellar populations with reasonable ages and thus,
provide a fairly good match with the observed stellar age-mass
relations (black lines) from \citet{Thomas05} and \citet{Thomas10}.   

\section{Summary and conclusion}\label{Discussion}

For this study, we have performed three sets of cosmological,
hydrodynamical zoom simulations of 45 halos with masses between
$10^{11} < M_{\mathrm{halo}} < 10^{13} M_\odot$ using a modified
Gadget2 version including metal enrichment and an empirical model for 
momentum driven galactic winds based on \citet{Oppenheimer06,
  Oppenheimer08}.  We separately (and differentially) study the
effects of metal enrichment/cooling and galactic winds on star
formation, cold gas fractions, assembly histories, size evolution,
metallicities and kinematic scaling relations of the central and
satellite galaxies. The main results can be summarised as follows:         

\begin{itemize}
\item[{\bf(i)}] Galactic winds can efficiently suppress early star
  formation in low and high mass galaxies as the gas inflow is delayed
  and gas is prevented to form stars by temporarily being blown out of
  the galaxies. This leads to an overall increase of cold gas fractions  
  and to a reduction of the baryon conversion efficiencies (for all
  halos at high redshifts and, in particular, for low mass halos 
  at low redshift). For galaxies in massive halos $M_{\mathrm{halo}}
  >  10^{12}$ galactic winds shift the peak in the star formation
  rate from z $\sim$ 5 to z $\sim 2$. For galaxies in lower mass halos
  the effect is even more dramatic shifting the peak from z $\sim$ 4
  to almost the present day. The resulting gas fractions, star
  formation rates and conversion efficiencies are in fairly good
  agreement with observations and theoretical estimates. Galactic
  winds provide a reasonable explanation for the observed trend that
  the star formation rates of low-mass galaxies peak at later cosmic
  times than the ones of massive galaxies (often termed as
  \textit{``stellar downsizing''}), i.e. low-mass galaxies form later
  in time and do not necessarily follow the early formation of their
  low-mass halos (as predicted by a cold dark matter dominated
  Universe). However, the present-day mean star formation rates and
  baryon conversion efficiencies of our most massive galaxies are by a
  factor of 10 and 2-3, respectively, too large. This clearly
  indicates that our currently adopted wind model is not sufficient
  for massive galaxies, possibly due to missing AGN feedback.  

\item[{\bf(ii)}] Galactic winds in combination with metal enrichment
  lead to an increasing importance of in situ star formation over stellar
  accretion for the stellar mass assembly over the entire redshift
  range. This is caused by more efficient metal cooling (more in situ
  star formation) and the lower galaxy masses in accreted satellite
  halos (less accretion of stars). The trend for in situ fractions to
  increase towards lower galaxy masses is in qualitative agreement
  with the estimates from abundance matching models
  (\citealp{Moster12, Behroozi12, Yang13}). Quantitatively, however,
  the WM models still predict too low in situ-to-accreted fractions
  for halos with masses below $M_{\mathrm{halo}} < 10^{12}
  M_\odot$. This is most likely due to a too large amount of accreted
  stellar systems, which could be suppressed by a more efficient 
  feedback for low-mass galaxies (e.g. \citealp{Dave13}). 

\item[{\bf(iii)}] Models with galactic winds under-estimate the sizes
  (effective radii) of present-day massive, quiescent galaxies
  ($M_{\mathrm{stellar}} > 10^{11} M_{\odot} $) as a consequence of
  the reduced fraction of accreted stellar systems and a too high in
  situ star formation at low redshifts. Models which additionally
  include AGN feedback may again be able to reproduce the observed,
  larger sizes. In contrast, the effective radii of lower mass,
  star-forming galaxies are strongly increased in the WM models in
  reasonably good agreement with the observed evolution. This holds
  true also for higher redshifts with the result that the relation
  between stellar mass and size in the wind model tends to vanish. As
  a consequence, the WM model predicts a much weaker size 
  evolution than the NoWM model which is in good, qualitative
  agreement for intermediate mass galaxies compared to  recent
  observational results. Overall, one of our main results is that none
  of our models is able to match the size  evolution for high-mass and
  low-mass galaxies, simultaneously.    

  In addition, the WM model (as opposed to models without winds) can
  successfully reproduce the observed relation between maximum
  circular velocity and stellar mass for galaxies with stellar masses
  below $M_{\mathrm{stellar}}<10^{11}M_{\odot}$ at $z=0,1,2$. At higher
  stellar masses the maximum circular velocities tend to too large in
  all models again pointing towards missing processes in the models as
  AGN feedback.  

\item[{\bf(iv)}] Galactic winds in zoom simulations are also found to
  delay the metal enrichment (as a consequence of the delayed star
  formation) and thus, to reduce the metallicity content (in both the
  stellar and gaseous component) resulting in an overall good match
  with observational data up to $z=2$. 
\end{itemize}
Our simulations complement previous studies by e.g. \citet{Finlator08,
  Dave09, Oppenheimer10, Dave11a} and \citet{Dave11b} (who use the
same model) by achieving a significantly increased mass and spatial
resolution. In addition, we extend the studies of \citet{Oser10} and
\citet{Oser12} whose simulations are based on the same initial
conditions but missing metal enrichment and strong stellar feedback. 

Overall, metal enrichment and strong galactic winds (even if
implemented in a very phenomenological way) are shown to be essential
for producing (particularly low-mass) galaxies with many reasonable
properties in excellent agreement with observations (with previous,
lower resolution studies). For massive galaxies at low redshifts,
however, some problems still remain: the baryon conversion
efficiencies and the in situ star formation rates are too high while
sizes are too small. This clearly signals one or more missing further
mechanisms (e.g. AGN feedback), to suppress late in situ star
formation and this way to increase the fraction of accreted stars.
Due to the possibly deficient galactic feedback (where the small-scale
physical processes driving the winds are not resolved) the necessity
for AGN feedback might be over-estimated. However, all of the
currently used, different models -- either simulations or
semi-analytic predictions -- agree that particularly for massive,
quiescent galaxies, another form of feedback is required, otherwise
the well-known over-cooling problem is encountered. A plausible
candidate  (as it was already shown in simulations and  SAMs) is
feedback from AGN. The relative strength of AGN and stellar feedback
is, however, still model dependent.
Moreover, some of our results (e.g. the too large contribution
of accreted stellar systems to the overall stellar mass assembly for
low-mass halos) indeed point towards a deficient implementation of
stellar feedback and they tend to indicate the need for even stronger
stellar winds particularly in low-mass galaxies. This is e.g. shown in
a very recent study of \citet{Dave13} who adopt an energy-driven wind
model for dwarf galaxies (where the mass loading scales with
$\sigma^{-2}$). This is found to even more reduce the amount low-mass  
galaxies and thus, match e.g. the low-mass end of the stellar mass
function better. In this paper, we have only focused on the effect of
metal enrichment and galactic winds on several global properties of
galaxies, but in a forthcoming paper, we additionally plan to study
the \textit{radial profiles} of different baryonic components,
particularly metallicity gradients, under the influence of galactic
winds and how they compare to recent observations. 

\section*{Acknowledgements}
MH acknowledges financial support from the European Research Council 
under the European Community's Seventh Framework Program
(FP7/2007-2013)/ERC grant agreement n. 202781.  
TN acknowledges support from the DFG cluster of excellence 'Origin and
Structure of the Universe'. Computer resources for this project have
been provided by the Leibniz Supercomputing Centre under
the grant nr. pr32re. We thank the referee, Fabio Governato, for a
careful and constructive reading of our paper. 

\bibliographystyle{mn2e}
\bibliography{Literaturdatenbank}

\label{lastpage}

\begin{appendix}
\section{Effects of numerical resolution}\label{resolution}
\begin{figure}
\begin{center}
  \epsfig{file=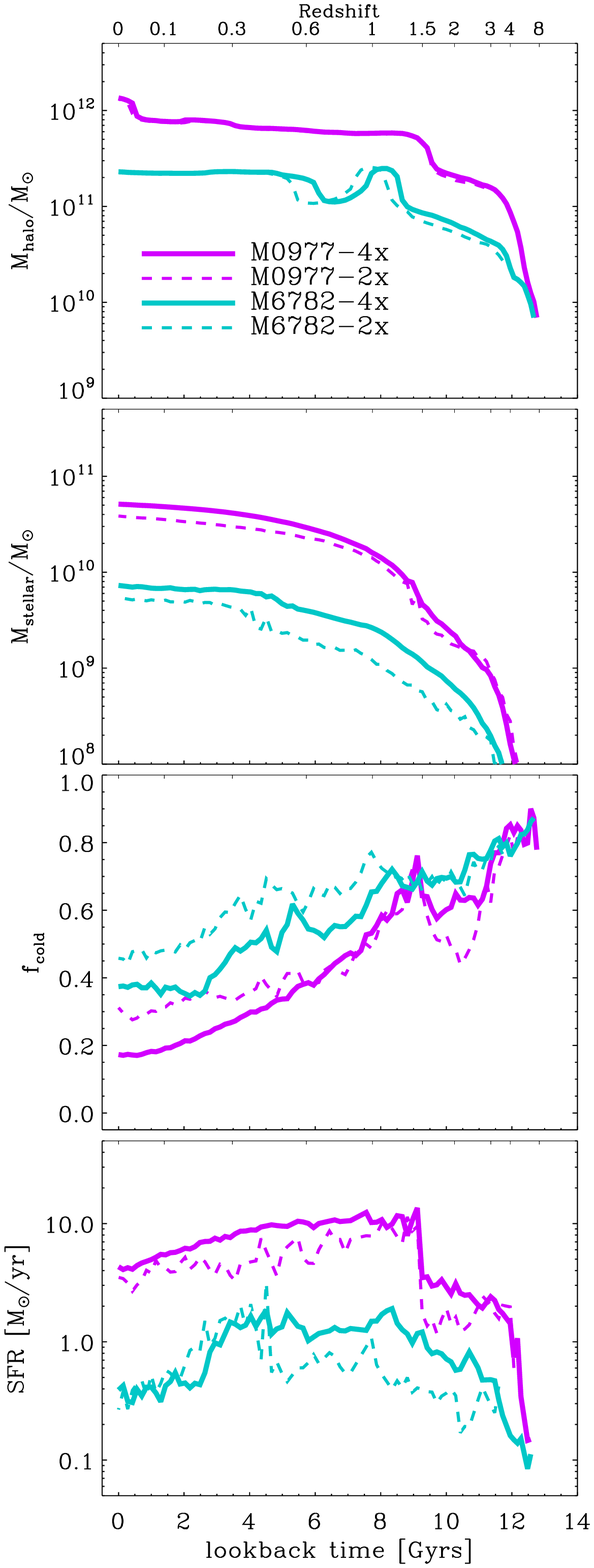, width=0.4\textwidth}
  \caption{Cosmic evolution of halo mass, stellar mass, cold gas
    fractions and SFRs for two different halos (M0977:
    $M_{\mathrm{halo}} \sim 10^{12} M_\odot$,  pink lines and
    M6782: $M_{\mathrm{halo}} \sim 2\times 10^{11} M_\odot$ turquoise
    lines) with 2x (dashed lines) and 4x (solid lines)
    resolution. We find no significant difference in the evolution of
    the central galaxy/main halo in the corresponding runs with
    different resolution.} 
  {\label{Comp_2x4x}} 
\end{center}
\end{figure}

For one high- and low-mass halo with final masses of
$M_{\mathrm{halo}} \sim 10^{12} M_\odot$ (M0977) and
$M_{\mathrm{halo}} \sim 2 \times 10^{11} M_\odot$ (M6782) we have
performed re-simulations including metal cooling and winds with a
\textit{four} times better spatial resolution (i.e. reduced softening
length by a factor of four, i.e. 4x-resimulations) than the original
one to study the convergence of our results with higher resolution. As
for the 2x-resolution runs, we traced back the particles that are
closer than $2 \times r_{200}$ to the centre of the halo in any of our
snapshots in the original DM-only  simulation ($m_{\mathrm{dm}} = 2
\times 10^8 M_\odot\ h^{-1}$) and replace them with dark matter and
gas particles of higher resolution, achieving a 64 times better mass
resolution in the high-resolution region than in the original DM
simulation (and thus, an eight times better mass resolution than the
2x-resimulations which are considered throughout this work):  
$m_{\mathrm{dm}} = 3.1 \times 10^6 M_\odot\ h^{-1}$ and 
$m_{\mathrm{gas}} = m_{\mathrm{star}} = 5.3 \times 10^5 M_\odot\
h^{-1}$.  

As shown in the upper left panel of Fig \ref{resolution} The mass
aggregation history of the dark matter component in both haloes (pink 
and turquoise lines) is very similar for the two different resolution
limits (4x-resolution: solid lines and 2x-resolution: dashed
lines). This also suggests that the haloes with different resolution
have experienced the same amount major/minor mergers. Regarding the
baryons, we show in Fig. \ref{resolution} the evolution of the stellar
mass, the cold gas fractions and the SFRs of the central galaxy. We
find that the haloes with four times larger spatial resolution contain
slightly more massive galaxies with slightly smaller cold gas
fractions, but higher SFRs. This indicates that with increasing
resolution the wind feedback seems to become very slightly less
efficient. This is only a weak effect, which we have already mentioned
before in order to explain the small differences between our
2x-resimulations and the results from previous studies using the same
code but having a worse resolution (\citealp{Oppenheimer06,
  Oppenheimer08, Dave09, Dave11a}). To summarise, we find no 
significant difference in the evolution of the central galaxy/main
halo in the simulations with different resolution. Therefore, due to
computational costs, we restrict our main study to 2x-resimulations
(i.e. the dashed lines in Fig. \ref{resolution}).

\end{appendix}

\end{document}